\pdfoutput=1
\documentclass[12pt,letterpaper]{article}
\usepackage[margin=1in]{geometry}
\usepackage[T1]{fontenc}
\usepackage[utf8]{inputenc}
\usepackage{lmodern}
\usepackage{setspace}
\usepackage{amsmath,amssymb,amsthm,bm}
\usepackage{booktabs}
\usepackage{graphicx}
\usepackage{placeins}
\usepackage{xcolor}
\usepackage[round,authoryear]{natbib}
\usepackage[hidelinks]{hyperref}
\usepackage[capitalise,nameinlink,noabbrev]{cleveref}

\DeclareMathOperator{\E}{\mathbb E}
\DeclareMathOperator{\Var}{Var}
\DeclareMathOperator{\Cov}{Cov}
\DeclareMathOperator{\tr}{tr}
\DeclareMathOperator{\diag}{diag}
\DeclareMathOperator{\KL}{KL}
\newcommand{\T}{\mathsf T}
\newcommand{\R}{\mathbb R}
\newcommand{\1}{\bm 1}
\newcommand{\given}{\,|\,}
\newcommand{\Normal}{\mathcal N}
\newtheorem{proposition}{Proposition}

\newcommand{\ReplicationPackage}{The replication package at
\url{https://github.com/Ruglio/PanelGP}}

\title{PanelGP: A Probabilistic Factor Model\\for Asset-Return Panels}
\author{%
\parbox[t]{0.45\textwidth}{\centering Andrea Ruglioni\\[0.5ex]
\normalsize EPFL, Lausanne, Switzerland\\
\href{mailto:andrea.ruglioni@epfl.ch}{andrea.ruglioni@epfl.ch}}%
\hspace{0.04\textwidth}%
\parbox[t]{0.45\textwidth}{\centering Damir Filipovic\\[0.5ex]
\normalsize EPFL, Lausanne, Switzerland\\
\href{mailto:damir.filipovic@epfl.ch}{damir.filipovic@epfl.ch}}}
\date{November 2026}

\hypersetup{
  pdftitle={PanelGP: A Probabilistic Factor Model for Asset-Return Panels},
  pdfauthor={Andrea Ruglioni and Damir Filipovic}
}

\begin{document}
\maketitle

\begin{abstract}
\noindent
We develop PanelGP, a conditional factor model for large panels of stock
returns in which firm characteristics determine the factor loadings and
economic conditions determine the expected factors. Both relations are
estimated nonparametrically with Gaussian-process priors. The kernel choice
nests instrumented principal component analysis (IPCA) as well as smooth
nonlinear maps.
Asset-specific loading innovations and common factor innovations separate
idiosyncratic from systematic risk, so one estimated model forecasts both
expected returns and their covariance matrix. Estimation uses sparse
variational inference with a joint posterior over loadings and factors.
Because each predicted return is bilinear in a Gaussian state, the
predictive covariance matrix and the cumulants of any portfolio have closed
forms. In simulations, the estimator recovers the nonlinear maps and the
variance parameters. For S\&P 500 constituents over 1995--2024, nonlinear
loadings rank stocks better than affine loadings and improve the Gaussian
log score. Relative to IPCA, the resulting portfolios have higher Sharpe
ratios, also net of trading costs, and the minimum-variance portfolio has
lower volatility. The predicted variances rank stocks and months by risk but
are too compressed.

\end{abstract}


\section{Introduction}
\label{sec:introduction}

A portfolio of individual stocks requires two forecasts for the same set of
stocks: expected returns, which determine which positions to take, and a
covariance matrix, which determines how much risk they carry. With hundreds
of stocks, neither can be estimated reliably from return histories alone. Firm
characteristics and economic conditions supply the missing information. A
large literature uses characteristics, often through nonlinear methods, to
forecast the cross-section of expected returns, and a separate literature
estimates large covariance matrices, typically with factor structures and
shrinkage.

A conditional factor model connects the two problems. In a factor model, the
expected return of a stock is the product of its loadings and the expected
factors, and the covariance between two stocks is the product of their
loadings and the factor covariance matrix. When the loadings are functions
of characteristics, one estimated map from characteristics to loadings
therefore delivers both forecasts. This paper develops PanelGP, a
conditional factor model in which both the map from characteristics to
loadings and the map from economic conditions, which we call the context, to
expected factors are estimated nonparametrically with Gaussian-process (GP)
priors. A single estimated model then gives expected returns, a covariance
matrix, and more generally a predictive distribution for the whole panel,
which also reflects posterior uncertainty about both maps.

Our first contribution is the model. Returns are bilinear in realized
loadings and realized factors. The realized loading of a stock is a smooth
function of its characteristics plus an asset-specific innovation, and the
realized factor is a smooth function of the context plus an innovation shared
by all stocks. The factor innovation creates covariance between stocks,
while the loading innovation adds only idiosyncratic variance. The kernels
determine the functional form. An affine loading kernel gives
the linear loadings of instrumented principal component analysis (IPCA;
\citealp{kelly_characteristics_2019}), a constant factor kernel gives a
factor forecast that ignores the context, and radial basis function (RBF)
kernels give smooth nonlinear maps. Comparing kernels within the same
estimator therefore isolates the effect of nonlinearity.

Our second contribution is inference and prediction. A sparse variational
approximation keeps a joint Gaussian posterior over inducing loadings and
realized training factors, with a dense covariance matrix, so the dependence
between loadings and factors is retained. Gaussian conditioning then gives
the loadings and factors at new characteristics and contexts, and each
predicted return is a quadratic form in this Gaussian state. Classical
results on Gaussian quadratic forms \citep{mathai_quadratic_1992} yield the
joint cumulant generating function of a new cross-section of returns, from
which the predictive covariance matrix and the cumulants of any portfolio
follow in closed form.

Our third contribution is empirical. In a correctly specified simulation,
the estimator recovers the smooth maps and the variance parameters. The
application uses historical S\&P 500 constituents from 1995 to 2024,
137 characteristics, and six macro-financial context
variables. The evidence is clearest for the ordering of stocks and for the
covariance matrix. Replacing affine by nonlinear loadings raises the rank
correlation between predicted and realized returns and the Gaussian log
score, and these differences survive a correction for multiple comparisons.
With five factors, the model with RBF loadings and RBF factors has a
significantly higher tangency Sharpe ratio and a significantly lower
minimum-variance volatility than IPCA with and without an intercept. Under
annual re-estimation, its tangency Sharpe ratio is 1.90 against
1.74 for rolling IPCA with an intercept; the tests
cannot separate these Sharpe ratios, but they do separate the log scores and
the minimum-variance volatilities. The model is less successful with the
level of individual variances: its predicted variances rank stocks and
months by risk but are too compressed.

\Cref{sec:related_work} reviews related work. \Cref{sec:model} specifies the
model, and \cref{sec:inference} derives the estimator and the predictive
distribution. \Cref{sec:synthetic} reports the simulation and
\cref{sec:asset_pricing} the application; \cref{sec:conclusion} concludes.
\Cref{app:variational,app:quadratic} contain the derivations,
\cref{app:specification} the estimation details, and
\cref{app:additional_results} additional results.

\section{Related work}
\label{sec:related_work}

Factor models summarize common return variation through a few factors and
asset-specific loadings
\citep{ross_arbitrage_1976,fama_common_1993,bai_inferential_2003}. Conditional
specifications relate the loadings to firm characteristics, through the
linear map of IPCA \citep{kelly_characteristics_2019}, through sieve
approximations of nonparametric loading functions
\citep{connor_efficient_2012,fan_projected_2016,chen_semiparametric_2023},
or through regularized high-dimensional estimation \citep{chen_unified_2026};
others let them vary with an economic state
\citep{lettau_factors_2020,pelger_state_2020}; see
\citet{giglio_factor_models_2022} for a review. Shrinkage and machine-learning
methods accommodate large characteristic sets and nonlinear relations
between characteristics and expected returns
\citep{freyberger_dissecting_2020,gu_empirical_2020,kelly_virtue_2024}, and
conditional autoencoders learn nonlinear loadings
\citep{gu_autoencoder_2021}. GP regression has been used for expected-return
prediction with uncertainty quantification \citep{filipovic_ensemble_2022}.

Most of this work targets expected returns. Using characteristics for both
moments goes back to the risk model of \citet{rosenberg_extra_1974}, whose
loadings on extra-market factors are functions of firm characteristics. IPCA
uses the same loadings for expected returns and covariances
\citep{kelly_characteristics_2019}, \citet{filipovic_cmk_2024} estimate the
conditional mean and covariance of an unbalanced panel jointly with one
kernel of characteristics, and \citet{filipovic_linear_2024} show that the
conditional mean-variance efficient portfolio of such a panel is spanned by
low-dimensional factor portfolios. A related literature bypasses the two
moments and estimates the stochastic discount factor, or equivalently the
mean-variance efficient portfolio, directly from characteristics, with
shrinkage
\citep{kozak_shrinking_2020}, cross-sectional regressions on characteristics
\citep{kozak_span_2023}, sparse regression
\citep{ao_approaching_2019}, asset-pricing trees
\citep{bryzgalova_forest_2025}, or neural networks and large factor models
\citep{chen_deep_2024,didisheim_aipt_2024,kelly_aipm_2025}. Parametric
portfolio policies model portfolio weights directly as functions of
characteristics \citep{brandt_parametric_2009,demiguel_transaction_2020},
and principal portfolios exploit cross-predictability between assets
\citep{kelly_principal_2023}. PanelGP instead produces a predictive
distribution for the panel, from which mean-variance portfolios are one of
several outputs.

Large covariance matrices of returns are usually estimated by shrinkage
\citep{ledoit_wolf_2004}, by imposing a factor structure
\citep{fan_factor_2008,fan_poet_2013}, or by combining both with dynamic
correlation models \citep{engle_large_2019}, and they are commonly evaluated
by the out-of-sample volatility of the global minimum-variance portfolio
\citep{ledoit_wolf_2017,engle_large_2019}. These estimators take expected
returns as given or ignore them. In PanelGP, posterior uncertainty about the
loadings and factors also enters the predictive covariance, as parameter
uncertainty does in Bayesian portfolio analysis \citep{avramov_bayesian_2010}.

GP priors specify distributions over functions through covariance kernels
\citep{rasmussen_gaussian_2006,alvarez_kernels_2012}. With one learned
lengthscale per coordinate, standard kernels remain effective in hundreds of
input dimensions \citep{binois_survey_2022,xu_standard_2025}. The closest
models are GP regression networks, which multiply GP weights and GP signals
to obtain input-dependent correlations \citep{wilson_gprn_2012}, and GP
models of input-dependent covariance matrices
\citep{wilson_wishart_2011,fox_dunson_2015}. PanelGP uses the bilinear
construction with separate inputs for assets and dates, allows an unbalanced
panel, and adds a factor innovation common to all assets at a date.
Variational inducing variables make estimation scalable
\citep{titsias_variational_2009,hensman_gaussian_2013,hoffman_stochastic_2013,bauer_understanding_2016}.
Variational approximations can understate predictive variance
\citep{blei_variational_2017,jankowiak_parametric_2020}, which is why we
check both the recovery of the latent functions and the calibration of the
predicted variances.

\section{The PanelGP model}
\label{sec:model}

\subsection{Panel factor structure}
\label{sec:factor_structure}

Input dates are indexed by $t=1,\ldots,T$. At date $t$, $\mathcal I_t$ is the
set of assets with characteristics observed at $t$ and an excess return
observed from $t$ to $t+1$, and $N_t=|\mathcal I_t|$, so the panel can be
unbalanced. Asset $i\in\mathcal I_t$ has characteristics
$\bm x_{it}\in\R^P$, and the context $\bm c_t\in\R^Q$ is common to all
assets.

The model has $H$ factors and two smooth maps. The loading map
$\bm\beta:\R^P\to\R^H$ maps characteristics to factor loadings, and the
factor map $\bm\phi:\R^Q\to\R^H$ maps the context to the expected factors of
the next period. Realized loadings and factors, marked by a tilde, deviate
from these maps by innovations:
\begin{align}
 \widetilde{\bm\beta}_{it}
 &=\bm\beta(\bm x_{it})+\bm\xi_{i,t+1},
 &\bm\xi_{i,t+1}&\stackrel{\mathrm{iid}}{\sim}\Normal(\bm0,\bm D_\beta),
 \label{eq:realized_loading}\\
 \widetilde{\bm\phi}_t
 &=\bm\phi(\bm c_t)+\bm\zeta_{t+1},
 &\bm\zeta_{t+1}&\stackrel{\mathrm{iid}}{\sim}\Normal(\bm0,\bm D_\phi).
 \label{eq:realized_factor}
\end{align}
Realized loadings and factors carry the input date $t$, while their
innovations realize between $t$ and $t+1$. The loading innovation
$\bm\xi_{i,t+1}$ is specific to an asset and a date, and the factor
innovation $\bm\zeta_{t+1}$ is shared by all assets at a date. The covariance
matrices $\bm D_\beta$ and $\bm D_\phi$ are diagonal with positive entries.
The excess return is
\begin{equation}
 r_{i,t+1}
 =\widetilde{\bm\beta}_{it}^{\T}\widetilde{\bm\phi}_t
 +\sigma_r\epsilon_{i,t+1},
 \qquad
 \epsilon_{i,t+1}\stackrel{\mathrm{iid}}{\sim}\Normal(0,1),
 \label{eq:model}
\end{equation}
where $\sigma_r\epsilon_{i,t+1}$ is the observation noise. The two
innovations and the observation noise are mutually independent and
independent of the inputs and the smooth maps. Loading innovations and
observation noise are independent across asset--date pairs; factor
innovations are independent across dates. Whatever the form of $\bm\beta$
and $\bm\phi$, the return is linear in the loadings given the factors and
linear in the factors given the loadings.

The training sample has $N=\sum_{t=1}^TN_t$ rows, indexed by
$n=1,\ldots,N$. Row $n$ refers to asset $i(n)$ at input date $t(n)$. We write
$r_n=r_{i(n),t(n)+1}$ and $\bm x_n=\bm x_{i(n)t(n)}$, and stack
\begin{equation}
 \bm r=\begin{bmatrix}r_1\\[-0.2em]\vdots\\[-0.2em]r_N\end{bmatrix},\qquad
 \bm X=
 \begin{bmatrix}\bm x_1^{\T}\\[-0.2em]\vdots\\\bm x_N^{\T}\end{bmatrix},
 \qquad
 \bm C=
 \begin{bmatrix}\bm c_1^{\T}\\[-0.2em]\vdots\\\bm c_T^{\T}\end{bmatrix},
 \qquad
 \bm B=
 \begin{bmatrix}\bm\beta(\bm x_1)^{\T}\\[-0.2em]\vdots\\
 \bm\beta(\bm x_N)^{\T}\end{bmatrix},
 \qquad
 \widetilde{\bm F}=
 \begin{bmatrix}\widetilde{\bm\phi}_1^{\T}\\[-0.2em]\vdots\\
 \widetilde{\bm\phi}_T^{\T}\end{bmatrix}.
 \label{eq:panel_matrices}
\end{equation}
Note that $\bm B$ holds smooth loadings while $\widetilde{\bm F}$ holds
realized factors; \cref{sec:likelihood} explains why.

\subsection{Moments and likelihood}
\label{sec:likelihood}

Conditional on the smooth maps, integrating out the innovations and the
observation noise gives the following moments. For $i\ne j$ at the same date,
\begin{equation}
 \Cov(r_{i,t+1},r_{j,t+1}\mid\bm\beta,\bm\phi)
 =\bm\beta(\bm x_{it})^{\T}\bm D_\phi
 \bm\beta(\bm x_{jt}).
 \label{eq:factor_innovation_covariance}
\end{equation}
For a single return,
\begin{align}
 \E(r_{i,t+1}\mid\bm\beta,\bm\phi)
 &=\bm\beta(\bm x_{it})^{\T}\bm\phi(\bm c_t),\notag\\
 \Var(r_{i,t+1}\mid\bm\beta,\bm\phi)
 &=\underbrace{\bm\beta(\bm x_{it})^{\T}\bm D_\phi\bm\beta(\bm x_{it})}
   _{\text{systematic}}
 +\underbrace{\bm\phi(\bm c_t)^{\T}\bm D_\beta\bm\phi(\bm c_t)
 +\tr(\bm D_\beta \bm D_\phi)+\sigma_r^2}_{\text{idiosyncratic}}.
 \label{eq:smooth_conditional_variance}
\end{align}
The systematic term comes from the factor innovation and is the only term
that also enters the covariances in \cref{eq:factor_innovation_covariance}.
The three idiosyncratic terms do not depend on the asset. Since
each innovation multiplies a smooth map, the variance depends on
$\bm x_{it}$ and $\bm c_t$ even though $\bm D_\beta$, $\bm D_\phi$, and
$\sigma_r^2$ are constant. \Cref{sec:prediction} gives the corresponding
moments of the estimated model, which also reflect posterior uncertainty
about $\bm\beta$ and $\bm\phi$.

For inference, we keep the realized factor and integrate out the loading
innovation:
\begin{equation}
 r_{i,t+1}\mid\bm\beta(\bm x_{it}),\widetilde{\bm\phi}_t
 \sim\Normal\!\left(
 \bm\beta(\bm x_{it})^{\T}\widetilde{\bm\phi}_t,\;
 \sigma_r^2+
 \widetilde{\bm\phi}_t^{\T}\bm D_\beta\widetilde{\bm\phi}_t
 \right).
 \label{eq:marginal_likelihood}
\end{equation}
Given $\bm B$ and $\widetilde{\bm F}$, the remaining loading innovations and
the observation noise are independent across rows, so the likelihood
factorizes:
\begin{equation}
 p(\bm r\mid\bm B,\widetilde{\bm F})
 =\prod_{n=1}^{N}
 p\big(r_n\mid\bm\beta(\bm x_n),\widetilde{\bm\phi}_{t(n)}\big).
 \label{eq:likelihood}
\end{equation}
Conditioning on the smooth factors $\bm\phi(\bm c_t)$ instead would leave
the factor innovation common to all returns of a date, and the likelihood
would no longer factorize. This is why the latent state contains realized
factors but smooth loadings.

\subsection{Gaussian-process priors}
\label{sec:priors}

The smooth maps have independent zero-mean GP priors,
\begin{equation}
 \beta_h\sim\mathcal{GP}(0,k_{\beta,h}),
 \qquad
 \phi_h\sim\mathcal{GP}(0,k_{\phi,h}),
 \qquad h=1,\ldots,H.
 \label{eq:gp_priors}
\end{equation}
At the training inputs, these priors imply
\begin{align}
 \bm B_{\cdot h}
 &\sim\Normal\!\left(\bm0,\bm K_{\beta,h}(\bm X,\bm X)\right),
 \label{eq:loading_gp_observed}\\
 \widetilde{\bm F}_{\cdot h}
 &\sim\Normal\!\left(
 \bm0,\bm K_{\phi,h}(\bm C,\bm C)
 +[\bm D_\phi]_{hh}\bm I_T\right),
 \label{eq:factor_gp_prior}
\end{align}
where $\bm B_{\cdot h}$ and $\widetilde{\bm F}_{\cdot h}$ are the $h$th
columns and the kernel matrices have entries
$k_{\beta,h}(\bm x_n,\bm x_{n'})$ and $k_{\phi,h}(\bm c_t,\bm c_{t'})$. The
loading innovation is absent from \cref{eq:loading_gp_observed} because
$\bm B$ holds smooth loadings; it enters through
\cref{eq:marginal_likelihood}.

Any positive definite kernels can be used. We build ours from three
elementary kernels. The affine kernel $1+\bm x^{\T}\bm x'$ generates affine
functions. The RBF kernel $\exp[-\frac12\sum_{p}(x_p-x'_p)^2/\ell_p^2]$,
with one lengthscale $\ell_p$ per input coordinate (automatic relevance
determination, ARD), generates smooth nonlinear functions. The constant
kernel $1$ generates functions that do not depend on the input.

We consider loading kernels of the form $k_{\beta,h}=\eta_h\bar k_{\beta,h}$,
where $\eta_h>0$ is an output scale and $\bar k_{\beta,h}$ is the affine or
the RBF kernel, and factor kernels of the form
\begin{equation}
 k_{\phi,h}(\bm c,\bm c')=a_h\,\bar k_{\mathrm{RBF}}(\bm c,\bm c';\bm\ell_{\phi,h})+b_h,
 \qquad a_h,b_h\ge0,\quad a_h+b_h+[\bm D_\phi]_{hh}=1,
 \label{eq:factor_kernel}
\end{equation}
where $\bar k_{\mathrm{RBF}}(\cdot,\cdot;\bm\ell)$ is the RBF kernel with
lengthscales $\bm\ell$.
The constant component matters for extrapolation. Far from the training
contexts the RBF component vanishes, and the constant component keeps a
context-invariant level learned from the training factors. We call this the
RBF factor kernel when $a_h>0$ and the constant factor kernel when $a_h=0$.
The constraint fixes the prior variance of each realized factor at one, so
the output scale $\eta_h$ carries the scale of the returns. All kernel
parameters are specific to the component $h$, and
\cref{app:kernel_specification} gives their parameterization. We write
$\bm\theta$ for the kernel parameters, the diagonal entries of $\bm D_\beta$
and $\bm D_\phi$, and $\sigma_r^2$.

These kernels are chosen to nest IPCA \citep{kelly_characteristics_2019}.
With the affine loading kernel and the constant factor kernel, the loadings
are affine in the characteristics and the realized factors vary from date to
date around a common level, so the factor forecast ignores the context. This
is the structure of IPCA, up to the loading innovation and the priors. The
RBF kernels generalize it in two directions: the loading kernel allows
nonlinear loadings, and the factor kernel allows
factor forecasts that depend on the context. \Cref{sec:kernel_results}
discusses how the two estimators differ.

\section{Estimation and prediction}
\label{sec:inference}

\subsection{Variational inference}
\label{sec:variational_objective}

Stack the smooth training loadings and the realized factors as
\begin{equation}
 \bm\chi=\begin{bmatrix}\operatorname{vec}(\bm B)\\
 \operatorname{vec}(\widetilde{\bm F})\end{bmatrix}.
 \label{eq:latent_vector}
\end{equation}
The mean of the likelihood in \cref{eq:marginal_likelihood} is bilinear in
$\bm\beta(\bm x_n)$ and $\widetilde{\bm\phi}_t$, and its variance depends on
$\widetilde{\bm\phi}_t$. The posterior $p_{\bm\theta}(\bm\chi\mid\bm r)$ is
therefore not Gaussian, and the marginal likelihood $p_{\bm\theta}(\bm r)$ has
no closed form. We approximate the posterior by a density
$q_{\bm\psi}(\bm\chi)$ with variational parameters $\bm\psi$. For any such
density,
\begin{equation}
\begin{aligned}
 \log p_{\bm\theta}(\bm r)
 &=\mathcal L(\bm\theta,\bm\psi)
 +\KL\!\left(q_{\bm\psi}(\bm\chi)\|p_{\bm\theta}(\bm\chi\mid\bm r)\right),\\
 \mathcal L(\bm\theta,\bm\psi)
 &=\E_q[\log p_{\bm\theta}(\bm r\mid\bm\chi)]
 -\KL\!\left(q_{\bm\psi}(\bm\chi)\|p_{\bm\theta}(\bm\chi)\right).
\end{aligned}
 \label{eq:main_variational_bound}
\end{equation}
Since the Kullback--Leibler (KL) divergence is nonnegative, $\mathcal L$ is
a lower bound on $\log p_{\bm\theta}(\bm r)$, the evidence lower bound
(ELBO; \citealp{blei_variational_2017}). Maximizing $\mathcal L$ over
$\bm\psi$ minimizes the divergence from $q$ to the posterior, and maximizing
it also over $\bm\theta$ estimates the model parameters.

We take $q$ to be Gaussian. A Gaussian cannot represent posterior skewness or
multiple modes, and all predictive results below are conditional on this
approximation; \citet{pmlr-v31-nguyen13b} use mixtures of Gaussians instead.
The state $\bm\chi$ has $(N+T)H$ components, several hundred thousand in our
application, which rules out a dense covariance matrix. We therefore use
inducing variables. For each $h$, we introduce $M\ll N$ inducing inputs
$\bm Z_h\in\R^{M\times P}$ and define the inducing values
$\bm u_h=\beta_h(\bm Z_h)$ and $\bm U=[\bm u_1,\ldots,\bm u_H]$.
The inducing locations are optimized with the model parameters. The reduced
state is
\begin{equation}
 \bm v=\begin{bmatrix}
  \operatorname{vec}(\bm U)\\
  \operatorname{vec}(\widetilde{\bm F})\end{bmatrix},
 \label{eq:reduced_latent_vector}
\end{equation}
with dimension $d_v=(M+T)H$. The variational family is
\begin{equation}
 q(\bm B,\bm U,\widetilde{\bm F})
 =p_{\bm\theta}(\bm B\mid\bm U)\,q_{\bm\psi}(\bm v),
 \qquad
 q_{\bm\psi}(\bm v)=\Normal(\bm m_v,\bm S_v),
 \label{eq:q_joint}
\end{equation}
where $p_{\bm\theta}(\bm B\mid\bm U)$ is the GP prior conditional. Because
the covariance $\bm S_v$ is dense, $q$ keeps the dependence across components
and between loadings and factors. When the identity in
\cref{eq:main_variational_bound} is applied to the augmented state
$(\bm B,\bm U,\widetilde{\bm F})$, the prior conditional cancels from the KL
term \citep{titsias_variational_2009,matthews_sparse_2016}, and
\begin{equation}
 \mathcal L(\bm\theta,\bm\psi)
 =\sum_{n=1}^{N}\E_q\!\left[
 \log p_{\bm\theta}\big(r_n\mid\bm\beta(\bm x_n),
 \widetilde{\bm\phi}_{t(n)}\big)\right]
 -\KL\!\left(q_{\bm\psi}(\bm v)\|
 p_{\bm\theta}(\bm U)p_{\bm\theta}(\widetilde{\bm F})\right).
 \label{eq:elbo}
\end{equation}
The KL term in \cref{eq:elbo} has a closed form, and the likelihood term is
a sum over rows by the conditional independence in \cref{eq:likelihood}. In
each row, we
integrate out the loading exactly given the factor and sample only the
$H$-dimensional factor, which gives unbiased gradients for minibatch
optimization. \Cref{app:variational} gives the derivations and
\cref{app:computation} the computational details.

\subsection{Prediction of the latent state}
\label{sec:latent_prediction}

Consider a new panel of $J$ rows at $W$ new dates with contexts
$\bm c_{\star1},\ldots,\bm c_{\star W}$; the context values need not be
distinct. Row $j$ has characteristics $\bm x_{\star j}$ and new date
$t(j)\in\{1,\ldots,W\}$. The realized loadings
$\widetilde{\bm\beta}_{\star j}^{\T}$ form the rows of
$\widetilde{\bm B}_\star\in\R^{J\times H}$, and the realized factors
$\widetilde{\bm\phi}_{\star t}^{\T}$ form the rows of
$\widetilde{\bm F}_\star\in\R^{W\times H}$. Stack
\begin{equation}
 \bm\chi_\star=
 \begin{bmatrix}
  \operatorname{vec}(\widetilde{\bm B}_\star)\\
  \operatorname{vec}(\widetilde{\bm F}_\star)
 \end{bmatrix}.
 \label{eq:new_latent_vector}
\end{equation}
Under the prior, $\bm\chi_\star$ and $\bm v$ are jointly Gaussian with mean
zero. Let $\widetilde{\bm K}_{\star\star}$ and $\widetilde{\bm K}_{vv}$ be the
prior covariance matrices of $\bm\chi_\star$ and $\bm v$, and $\bm K_{\star v}$
their cross-covariance. The diagonal blocks of
$\widetilde{\bm K}_{\star\star}$ contain the covariances of the new
innovations, and $\widetilde{\bm K}_{vv}$ contains those of the training
factor innovations. Since the new innovations are independent of the
training state, the cross-covariance contains only smooth kernel terms.
Gaussian conditioning gives
\begin{equation}
 p_{\bm\theta}(\bm\chi_\star\mid\bm v)
 =
 \Normal(\bm G_\star\bm v,\bm R_\star),\qquad
 \bm G_\star=\bm K_{\star v}\widetilde{\bm K}_{vv}^{-1},\qquad
 \bm R_\star=\widetilde{\bm K}_{\star\star}
 -\bm K_{\star v}\widetilde{\bm K}_{vv}^{-1}\bm K_{\star v}^{\T}.
 \label{eq:latent_gaussian_conditioning}
\end{equation}
Integrating over $q_{\bm\psi}(\bm v)$ gives the predictive distribution of
the latent state,
\begin{equation}
 q(\bm\chi_\star)
 =
 \Normal(\bm m_\star,\widetilde{\bm S}_\star),
 \qquad
 \bm m_\star=\bm G_\star\bm m_v,\qquad
 \widetilde{\bm S}_\star=\bm R_\star+\bm G_\star \bm S_v\bm G_\star^{\T}.
 \label{eq:joint_latent_prediction}
\end{equation}
Here $\bm R_\star$ is the prior uncertainty that remains given $\bm v$, and
$\bm G_\star \bm S_v\bm G_\star^{\T}$ adds the posterior uncertainty about
$\bm v$, including the dependence between loadings and factors. Since
$\bm D_\beta$ and $\bm D_\phi$ are positive definite, so is
$\widetilde{\bm S}_\star$. \Cref{app:predictive_conditioning} gives the
blocks of $\bm m_\star$ and $\widetilde{\bm S}_\star$.

\subsection{Predictive distribution of returns}
\label{sec:prediction}

The return of new row $j$ is
\begin{equation}
 r_{\star j}
 =\widetilde{\bm\beta}_{\star j}^{\T}
 \widetilde{\bm\phi}_{\star t(j)}
 +\sigma_r\epsilon_{\star j}
 =\bm\chi_\star^{\T}\bm A_j\bm\chi_\star
 +\sigma_r\epsilon_{\star j},
 \label{eq:quadratic_form}
\end{equation}
where $\bm A_j$ is the symmetric matrix with entries $\frac12$ that pair the
coordinates of $\widetilde{\bm\beta}_{\star j}$ with those of
$\widetilde{\bm\phi}_{\star t(j)}$, and zeros elsewhere
(\cref{app:new_panel_law}). Let
$\bm r_\star=[r_{\star1},\ldots,r_{\star J}]^{\T}$, and for
$\bm s\in\R^J$ let $\bm A(\bm s)=\sum_{j=1}^Js_j\bm A_j$.

\begin{proposition}[Joint predictive CGF]
 \label{prop:joint_cgf}
Under $q(\bm\chi_\star)=\Normal(\bm m_\star,\widetilde{\bm S}_\star)$, the
cumulant generating function (CGF)
$\mathcal K(\bm s)=\log\E\exp(\bm s^{\T}\bm r_\star)$ is finite if and only if
$\widetilde{\bm S}_\star^{-1}-2\bm A(\bm s)$ is positive definite. On this set,
\begin{equation}
 \mathcal K(\bm s)
 =
 -\frac12\log|\bm I-2\widetilde{\bm S}_\star \bm A(\bm s)|
 +\bm m_\star^{\T}\bm A(\bm s)
 \left(\bm I-2\widetilde{\bm S}_\star \bm A(\bm s)\right)^{-1}
 \bm m_\star
 +\frac12\sigma_r^2\bm s^{\T}\bm s.
 \label{eq:cgf}
\end{equation}
\end{proposition}

\begin{proposition}[Cumulants]
\label{prop:all_cumulants}
Let $\bm w\in\R^J$ and $r_{\bm w}=\bm w^{\T}\bm r_\star$. For every integer
$k\ge1$, the $k$th cumulant of $r_{\bm w}$ is
\begin{equation}
 \kappa_k(r_{\bm w})
 =
 2^{k-1}(k-1)!\left[
 \tr\!\left((\bm A(\bm w)\widetilde{\bm S}_\star)^k\right)
 +k\,\bm m_\star^{\T}
 (\bm A(\bm w)\widetilde{\bm S}_\star)^{k-1}
 \bm A(\bm w)\bm m_\star
 \right]
 +
 \delta_{k2}\sigma_r^2\|\bm w\|_2^2,
 \label{eq:cumulants}
\end{equation}
where $\delta_{k2}$ is the Kronecker delta.
\end{proposition}

For portfolio weights $\bm w$, $r_{\bm w}$ is the portfolio return, and for
the $j$th unit vector $\bm w=\bm e_j$ it is the return of row $j$. Mixed
derivatives of $\mathcal K$ at zero give the cross-cumulants of the new
returns. \Cref{app:new_panel_law,app:scalar_functionals} prove
\cref{prop:joint_cgf,prop:all_cumulants}. Because $\mathcal K$ is finite near
zero, the tails of the predictive distribution decay at least exponentially:
the distribution can be skewed and have excess kurtosis, but it cannot have
power-law tails (\cref{app:scalar_functionals}).

Denote by $\bm m_{\beta,j}$ and $\bm m_{\phi,t}$ the blocks of $\bm m_\star$
for $\widetilde{\bm\beta}_{\star j}$ and $\widetilde{\bm\phi}_{\star t}$, and
by $[\widetilde{\bm S}_\star]_{\beta_j\phi_t}
=\Cov_q(\widetilde{\bm\beta}_{\star j},\widetilde{\bm\phi}_{\star t})$ and
similar symbols the blocks of $\widetilde{\bm S}_\star$. The first cumulant is
the predictive mean
\begin{equation}
\hat\mu_{\star j}
 =
 \bm m_{\beta,j}^{\T}\bm m_{\phi,t(j)}
 +\tr\!\left([\widetilde{\bm S}_\star]_{\beta_j\phi_{t(j)}}\right).
 \label{eq:predictive_mean}
\end{equation}
The trace term is the posterior covariance between the loading and the factor
of row $j$. The predictive covariance matrix
$\hat{\bm\Sigma}_\star=[\hat\Sigma_{\star,jj'}]$ follows from the second
derivatives of $\mathcal K$ and is given in \cref{eq:predictive_covariance}.
In \cref{sec:asset_pricing}, each new panel is the cross-section of a single
month $t$ ($W=1$), and we write $\hat\mu_{it}$ and $\hat{\bm\Sigma}_t$ for
the predictive mean of asset $i$ and the predictive covariance matrix.

To see where the predicted variance comes from, remove the new innovations
from the diagonal blocks of $\widetilde{\bm S}_\star$ and call the result
$\bm S_\star$:
\begin{equation}
 [\widetilde{\bm S}_\star]_{\beta_j\beta_j}
   =[\bm S_\star]_{\beta_j\beta_j}+\bm D_\beta,
 \qquad
 [\widetilde{\bm S}_\star]_{\phi_t\phi_t}
   =[\bm S_\star]_{\phi_t\phi_t}+\bm D_\phi,
 \label{eq:noisy_smooth_covariance_blocks}
\end{equation}
with all other blocks equal. The matrix $\bm S_\star$ is the posterior
covariance of the smooth values $\bm\beta(\bm x_{\star j})$ and
$\bm\phi(\bm c_{\star t})$. Substituting \cref{eq:noisy_smooth_covariance_blocks}
into \cref{eq:predictive_covariance} gives, for row $j$ at new date
$t=t(j)$,
\begin{equation}
 \Var_q(r_{\star j})
 =V^0_j+V^\phi_j+V^\beta_t+\tr(\bm D_\beta \bm D_\phi)+\sigma_r^2,
 \label{eq:innovation_variance_partition}
\end{equation}
where
$V^0_j=\Var_q\big(\bm\beta(\bm x_{\star j})^{\T}\bm\phi(\bm c_{\star t})\big)$
is the variance due to posterior uncertainty about the smooth maps, and
\begin{align}
 V^\phi_j
 &=\bm m_{\beta,j}^{\T}\bm D_\phi\bm m_{\beta,j}
   +\tr\!\left([\bm S_\star]_{\beta_j\beta_j}\bm D_\phi\right),\notag\\
 V^\beta_t
 &=\bm m_{\phi,t}^{\T}\bm D_\beta\bm m_{\phi,t}
   +\tr\!\left(\bm D_\beta[\bm S_\star]_{\phi_t\phi_t}\right).
 \label{eq:innovation_variance_terms}
\end{align}
The term $V^\phi_j$ is the variance added by the factor innovation,
$V^\beta_t$ the variance added by the loading innovation, and
$\tr(\bm D_\beta\bm D_\phi)$ the variance of the product of the two
innovations. At a given date, predicted variances differ across assets only
through $V^0_j$ and $V^\phi_j$, that is, through the loadings and their
posterior uncertainty.

\section{Simulation}
\label{sec:synthetic}

Before turning to data, we check whether the smooth maps, the realized latent
states, and the variance parameters can be recovered from returns that the
model itself generates.

\subsection{Design}
\label{sec:synthetic_design}

The balanced panel has $80$ assets and $50$ months, or
$4{,}000$ rows. Asset $i$ has a fixed characteristic vector $\bm x_i$ on a
regular $9\times9$ grid in $[0,1]^2$ with its upper-right corner omitted, and
month $t$ has the scalar context $c_t=(t-1)/(50-1)$. With $H=2$, the
smooth maps are
\begin{align}
 \phi_1(c)&=\tfrac12+\sin(2\pi c)+\tfrac12\cos(4\pi c)+\tfrac13\sin(6\pi c),
 \notag\\
 \phi_2(c)&=-\tfrac12+\cos(2\pi c)-\tfrac12\sin(4\pi c)+\tfrac13\cos(6\pi c),
 \notag\\
 \beta_1(\bm x)&=\sin(2\pi x_1)\cos(\pi x_2)
 +\tfrac12\cos\!\left(2\pi(x_1+x_2)\right)
 +\tfrac13\sin(4\pi x_1)\sin(2\pi x_2),
 \notag\\
 \beta_2(\bm x)&=\cos(2\pi x_2)\sin(\pi x_1)
 -\tfrac12\sin\!\left(2\pi(x_1-x_2)\right)
 +\tfrac13\cos(2\pi x_1)\cos(4\pi x_2).
 \label{eq:synthetic_functions}
\end{align}
Each function has several local extrema, so neither affine loadings nor a
constant factor map can reproduce it. Returns follow \cref{eq:model} with
$\bm D_\beta=\sigma_\beta^2\bm I$, $\bm D_\phi=\sigma_\phi^2\bm I$, and
$\sigma_\beta^2=\sigma_\phi^2=\sigma_r^2=0.05$.

We estimate the model on the full panel with $H=2$, RBF loading and factor
kernels (the factor kernel includes the constant term of
\cref{eq:factor_kernel}), 64 trainable inducing inputs, a dense $\bm S_v$, and
$5{,}000$ epochs. The estimator knows the number of factors and the
bilinear form, but not the smooth maps, the latent states, or the variances.
As in the design, it imposes isotropic innovations,
$\bm D_\beta=\sigma_\beta^2\bm I$ and $\bm D_\phi=\sigma_\phi^2\bm I$, while
the kernel parameters differ across the two components.

The factors are identified only up to an invertible linear transformation:
the product $\widetilde{\bm\beta}^{\T}\widetilde{\bm\phi}$ is unchanged when
the factors are replaced by $\bm\Upsilon\widetilde{\bm\phi}$ and the loadings
by $\bm\Upsilon^{-\T}\widetilde{\bm\beta}$ for a nonsingular $\bm\Upsilon$.
For evaluation, we align the estimated factors to the true ones by least
squares and transform the loadings by $\bm\Upsilon^{-\T}$. The alignment
also transforms the estimated innovation covariances, to
$\bm\Upsilon\hat{\bm D}_\phi\bm\Upsilon^{\T}$ and
$\bm\Upsilon^{-\T}\hat{\bm D}_\beta\bm\Upsilon^{-1}$, and
\cref{tab:synthetic_distribution} reports the average diagonal entry of each
transformed matrix. The return moments reported below do not depend on the
alignment. Loading and observation variances are expressed in original
return units (\cref{app:target_scale}). All results average five
replications, each with its own simulated panel and optimization seed.

\subsection{Results}
\label{sec:synthetic_functions_recovery}

\begin{figure}[!tb]
\centering
\includegraphics[width=\textwidth]{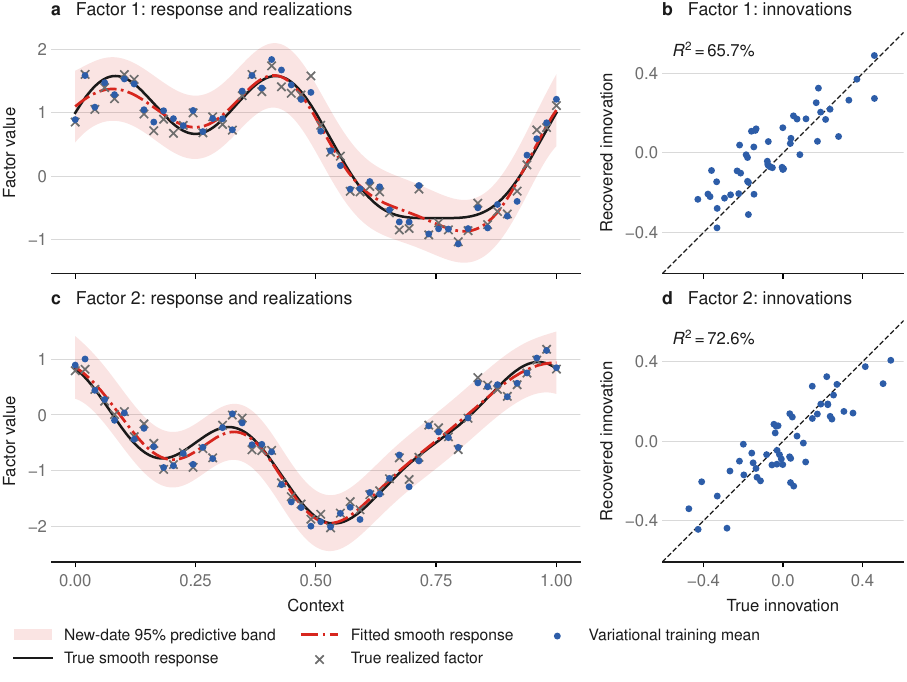}
\caption{Recovery of the smooth factor responses and the realized training
factors in replication~4{}. Each row shows one factor.
Left: the true smooth function (black), its fitted predictive mean
(dash-dotted red), the true realized factors (gray crosses), and their
variational posterior means (blue points). The shaded band is the fitted 95\%
interval for a new date, including a fresh innovation. Right: true
innovation (realized factor minus true smooth function) against recovered
innovation (variational mean minus fitted smooth function) at all
50 training dates; the diagonal marks perfect recovery, and the
reported $R^2$ measures innovation recovery. Both
columns use the alignment described in the text.}
\label{fig:synthetic_factors}
\end{figure}

\Cref{fig:synthetic_factors} shows the recovery of the factors in
replication~4{}, the one with the median average
factor-function $R^2$. It separates two objects. The predictive mean at a
new date follows the true smooth function, and its predictive interval
includes a fresh factor innovation. At the training dates, the variational
factors capture the smooth response and also much of the realized factor
innovation, with an $R^2$ of $65.7\%$ and
$72.6\%$ for the two factors, because the cross-section of
returns at that date is informative about it; a forecast for a new date has
no such information and integrates over the innovation. Across the five
replications, the fitted factor functions explain on average
$98.68\%$ of the variation of the true ones and the fitted loading functions $99.24\%$
(\cref{fig:synthetic_loadings}), with both $R^2$ values computed on dense
grids over the input domain.

The fitted predictive means explain $98.34\%$ of the variation of the
true conditional means $\bm\beta(\bm x_i)^{\T}\bm\phi(c_t)$. The fitted
predictive variances have a correlation of $0.98$ with
the oracle variances of \cref{eq:smooth_conditional_variance}, computed from
the true smooth maps, and the ratio of their averages is
$1.04$. The estimated innovation and noise variances are close
to their true value of $0.05$ (\cref{tab:synthetic_distribution}). When the
model is correctly specified, the estimator therefore recovers the smooth
maps, the conditional return moments, and the variance parameters.

\begin{table}[!tb]
\centering
\caption{Recovery in five replications. Function $R^2$ values are computed on
dense input grids. Innovation variances are reported after alignment to the
true factors. The standard deviation across replications reflects both the
simulated panel and the optimization seed.}
\label{tab:synthetic_distribution}
\begingroup\small
\setlength{\tabcolsep}{4pt}\renewcommand{\arraystretch}{1.15}
\begin{tabular*}{\textwidth}{@{\extracolsep{\fill}}lrrr@{}}
\toprule
Quantity & True & Estimated mean & Replication s.d. \\
\midrule
Factor function $R^2$ (\%) & 100.00 & 98.68 & 0.53 \\
Loading function $R^2$ (\%) & 100.00 & 99.24 & 0.19 \\
Loading innovation $\sigma_\beta^2$ & 0.050 & 0.048 & 0.001 \\
Factor innovation $\sigma_\phi^2$ & 0.050 & 0.049 & 0.009 \\
Observation noise $\sigma_r^2$ & 0.050 & 0.052 & 0.004 \\
\bottomrule
\end{tabular*}
\endgroup

\end{table}

\section{Application to S\&P 500 constituents}
\label{sec:asset_pricing}

We now apply the model to monthly returns of S\&P 500 constituents and
evaluate its mean and variance forecasts and the portfolios built from them.

\subsection{Data and estimation design}
\label{sec:empirical_design}

Monthly U.S. stock characteristics and returns come from the global factor
data of \citet{jensen_is_2023}, distributed through WRDS. We use common stocks
on the main exchanges and one primary security per firm. Accounting
characteristics follow the source's convention of a four-month lag after the
fiscal period ends. Historical S\&P 500 membership
dates come from CRSP, and we keep the stock--months whose input month-end
falls within a period of index membership. Characteristics are selected by
availability: of the 153 characteristics of \citet{jensen_is_2023}, we use
the 137 that are observed for at least $70\%$ of these
stock--months in the 1960--1989 training period, and we keep the
stock--months for which at least $70\%$ of them are observed. The feature set
is then held fixed. Within each month, observed characteristics are ranked
and mapped to $[-0.5,0.5]$, and missing values are set to zero. The
evaluation panel, January 1995 to December 2024, has 173{,}740\
stock--months over 360 input months, with 465 to 497 stocks
per month.

The context consists of six variables from \citet{goyal_comprehensive_2024}:
the market return, the term spread, the default-yield spread, the log
realized market variance, the average correlation, and inflation, which we
lag by one month because of its publication delay. The characteristics and
context of month $t$ predict the excess return of month $t+1$, in decimal
monthly units. Before estimation, we standardize the context and scale
returns using training data only (\cref{app:target_scale}); forecasts are
transformed back before evaluation.

In the fixed-training design, the model is estimated on January 1960 to
December 1989, validated on 1990--1994, and evaluated on January 1995 to
December 2024. The parameters stay fixed over the evaluation period, but each
forecast uses the current characteristics and context. All PanelGP models use
160 trainable inducing inputs per component, a dense $\bm S_v$, ARD
lengthscales in every RBF kernel, and 500 training epochs.

We estimate each kernel pair for $H=1,\ldots,8$ with five optimization seeds,
holding epochs, inducing points, learning rates, and initialization fixed
across these runs. The validation results favor different dimensions for
different criteria, and we use $H=5$ as a compromise between validation
performance and model complexity (\cref{fig:validation}); the rolling
estimation uses the same dimension. The spread across seeds measures the
sensitivity of the results to the optimization.

\subsection{IPCA benchmarks}
\label{sec:kernel_ablation}

IPCA provides two benchmarks. Let $\bm X_t$ be the $N_t\times P$ matrix of
characteristics at month $t$. Both benchmarks have loadings
$[\bm X_t,\1]\bm\Gamma$ with $\bm\Gamma\in\R^{(P+1)\times H}$, the same affine
form as the affine PanelGP kernel. The second benchmark adds an intercept
$[\bm X_t,\1]\bm\gamma_\alpha$, estimated as the loading on a factor fixed at
one. The forecasts of the first benchmark are
\begin{equation}
 \hat{\bm\mu}^{\mathrm{IPCA}}_t=[\bm X_t,\1]\hat{\bm\Gamma}\bar{\bm\phi},\qquad
 \hat{\bm\Sigma}^{\mathrm{IPCA}}_t=
 [\bm X_t,\1]\hat{\bm\Gamma}\hat{\bm\Sigma}_\phi\hat{\bm\Gamma}^{\T}[\bm X_t,\1]^{\T}
 +\hat\sigma_e^2\bm I,
 \label{eq:ipca_predictive_moments}
\end{equation}
where $\bar{\bm\phi}$ and $\hat{\bm\Sigma}_\phi$ are the sample mean and
covariance of the estimated training factors, and $\hat\sigma_e^2$ is the
pooled in-sample residual variance. The intercept benchmark adds
$[\bm X_t,\1]\hat{\bm\gamma}_\alpha$ to the mean. Both are estimated by
alternating least squares.

\subsection{Evaluation criteria}
\label{sec:empirical_evaluation}

Let $\hat{\bm\mu}_t$ and $\hat{\bm\Sigma}_t$ be the predictive mean and
covariance matrix for the returns of month $t+1$. The predictive $R^2$,
\begin{equation}
 R^2_{\mathrm{pred}}=1-
 \frac{\sum_t\sum_{i\in\mathcal I_t}(r_{i,t+1}-\hat\mu_{it})^2}
      {\sum_t\sum_{i\in\mathcal I_t}r_{i,t+1}^2},
 \label{eq:empirical_r2}
\end{equation}
pools squared errors over the evaluation panel and compares them with those
of a zero forecast. The rank correlation is the monthly Spearman correlation
between predicted and realized returns, averaged over months; it measures
how well the forecasts order stocks. The Gaussian log score of month $t$ is
\begin{equation}
 \ell_t=-\frac{N_t\log(2\pi)+\log|\hat{\bm\Sigma}_t|
 +(\bm r_{t+1}-\hat{\bm\mu}_t)^{\T}\hat{\bm\Sigma}_t^{-1}
 (\bm r_{t+1}-\hat{\bm\mu}_t)}{2N_t},
 \label{eq:joint_nll}
\end{equation}
averaged over months. It evaluates the predictive mean and covariance matrix
jointly, but not the higher moments of the predictive distribution.

We form three portfolios each month. The global minimum-variance (GMV)
portfolio and the tangency portfolio are
\begin{equation}
 \bm w_t^{\mathrm{GMV}}
 =\frac{\hat{\bm\Sigma}_t^{-1}\1}
 {\1^{\T}\hat{\bm\Sigma}_t^{-1}\1},\qquad
 \bm w_t^{\mathrm{tan}}=\frac{\hat{\bm\Sigma}_t^{-1}\hat{\bm\mu}_t}
 {\|\hat{\bm\Sigma}_t^{-1}\hat{\bm\mu}_t\|_1}.
 \label{eq:portfolio_weights}
\end{equation}
Short positions are allowed. The GMV weights sum to one, and the tangency
weights have unit gross exposure. The long--short portfolio buys the decile
of stocks with the highest predicted returns and sells the decile with the
lowest, with equal weights and a gross exposure of one-half on each side, so
it uses the ordering of the forecasts but not the covariance matrix.

\subsection{Kernel comparison}
\label{sec:kernel_results}

\begin{table}[!tb]
\centering
\caption{Kernel pairs and IPCA benchmarks with $H=5$. PanelGP entries average
five seeds. Rank is the average monthly Spearman correlation and log score
the average monthly Gaussian log score. Tangency and Long--short are
annualized Sharpe ratios before trading costs, and GMV vol.\ is annualized
volatility. IPCA+int.\ adds an intercept to IPCA. Bold marks the best point
estimate in each column (lowest for GMV vol.); pairwise tests are in
\cref{tab:test_statistics}.}
\label{tab:focused_kernels}
\begingroup\small
\setlength{\tabcolsep}{4pt}\renewcommand{\arraystretch}{1.15}
\begin{tabular*}{\textwidth}{@{\extracolsep{\fill}}llrrrrrr@{}}
\toprule
 & & \multicolumn{3}{c}{Return forecasts} & \multicolumn{3}{c}{Portfolio performance} \\
\cmidrule(lr){3-5}\cmidrule(lr){6-8}
Loadings & Factors & $R^2$ (\%) & Rank (\%) & Log score & Tangency & Long--short & GMV vol.\ (\%) \\
\midrule
Affine & const. & 0.538 & 3.32 & 1.059 & 1.29 & 0.83 & \textbf{14.0} \\
Affine & RBF & 0.650 & 3.70 & 1.086 & 1.34 & 0.93 & 14.2 \\
RBF & const. & 0.542 & 3.97 & 1.067 & 1.41 & 0.88 & 14.7 \\
RBF & RBF & \textbf{0.753} & \textbf{4.52} & \textbf{1.092} & \textbf{1.55} & \textbf{0.98} & 14.3 \\
\addlinespace[0.5ex]
IPCA & -- & 0.310 & 2.59 & 1.066 & 0.84 & 0.57 & 15.9 \\
\addlinespace[0.5ex]
IPCA+int. & -- & 0.285 & 3.11 & 1.061 & 1.17 & 0.65 & 16.3 \\
\bottomrule
\end{tabular*}
\endgroup

\end{table}

Among the four PanelGP kernel pairs in \cref{tab:focused_kernels}, RBF
loadings with RBF factors give the highest predictive $R^2$, rank
correlation, Gaussian log score, and both portfolio Sharpe ratios. Only the
GMV volatility is lower elsewhere, under both pairs with affine loadings.
\Cref{tab:test_statistics} reports pairwise tests with a Holm adjustment for
multiple comparisons, and we call a difference significant when its adjusted
$p$-value is below $5\%$. Nonlinear loadings matter most: replacing affine by
RBF loadings raises the rank correlation and the log score significantly
under either factor kernel. A context-dependent factor forecast raises the
predictive $R^2$, the rank correlation, the log score, and both Sharpe ratios
under either loading kernel, but by amounts that the tests cannot
distinguish from zero. Against both IPCA benchmarks, RBF loadings with RBF
factors give a higher tangency Sharpe ratio and a lower GMV volatility, and
against IPCA without an intercept also a higher rank correlation, all
significantly. The tests detect no difference in pooled $R^2$ between
PanelGP kernel pairs. This criterion is sensitive to a few extreme returns
and market episodes, and we give more weight to ranks, log scores, and
portfolio outcomes.

Adding the intercept raises the tangency Sharpe ratio of IPCA from
$0.84$ to $1.17$. The affine PanelGP
with a constant factor kernel has the structure of IPCA (\cref{sec:priors}),
but the estimators differ. In the mean forecast, the GP prior shrinks the
affine coefficients, and the factor forecast is a shrunken average of the
realized factors rather than their sample mean. In the covariance matrix,
PanelGP adds the posterior uncertainty about loadings and factors to the
innovation variances.

\begin{figure}[!tb]
\centering
\includegraphics[width=\textwidth]{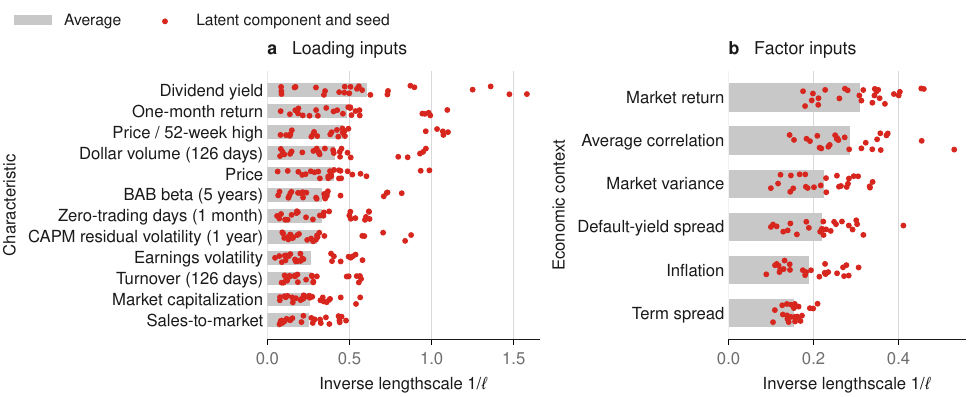}
\caption{Inverse ARD lengthscales of the fixed-training model with RBF
loadings, RBF factors, and $H=5$. Bars average over five components and five
seeds; points show the 25 component--seed values. The loading panel shows the
twelve characteristics with the largest averages, and the factor panel shows
all six context variables.}
\label{fig:ard_lengthscales}
\end{figure}

\Cref{fig:ard_lengthscales} reports the inverse ARD lengthscales
$1/\ell_{\beta,hp}$ and $1/\ell_{\phi,hq}$ of the model with RBF loadings and
RBF factors. A larger value means that the fitted map varies more quickly
along that input. Averaged over components and seeds, the largest values in
the loading kernel belong to dividend yield, the one-month return, price
relative to its 52-week high, dollar volume, and price, and those of the
factor kernel to the market return and the average correlation. Inputs share
a common scale within each panel, ranks for characteristics and standardized
values for the context, but the inverse lengthscales are not weighted by the
output scales. They indicate where the maps are most nonlinear rather than
how much each input contributes to the predicted returns. Since missing
ranks are set to zero, they may also reflect patterns of missing data.

\subsection{Number of factors}
\label{sec:hidden_dimension}

\begin{figure}[!tb]
\centering
\includegraphics[width=\textwidth]{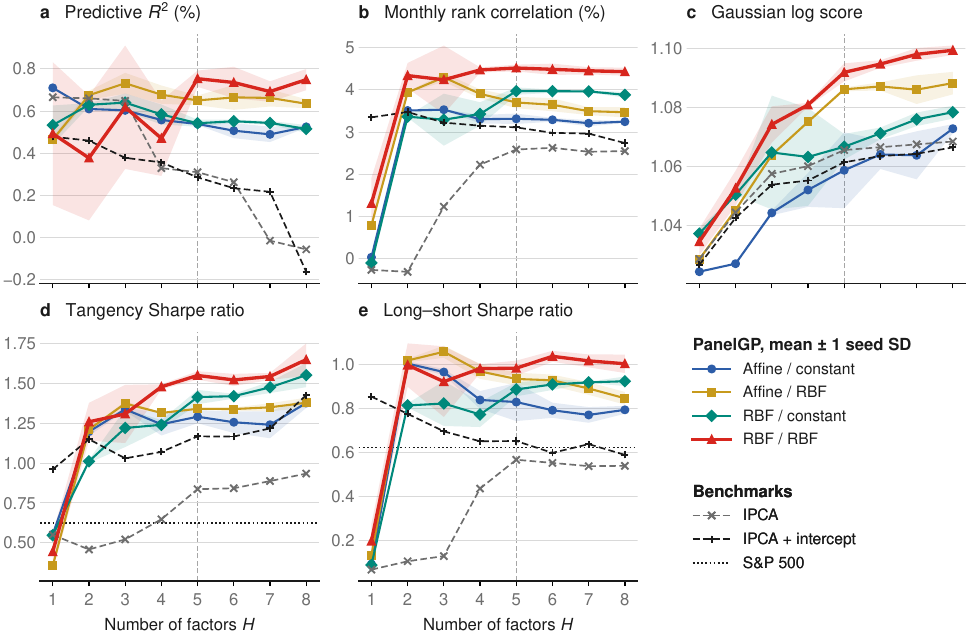}
\caption{Results by number of factors $H$ over the evaluation period.
PanelGP curves show five-seed means, and bands extend one sample standard
deviation across seeds on either side, measuring optimization dispersion.
IPCA uses a deterministic initialization. The S\&P 500 line is the
annualized Sharpe ratio of the index's total return minus the risk-free
rate over the same realization months as the portfolio returns. All kernel
pairs use the same data and training settings. Portfolio Sharpe ratios are
before trading costs.}
\label{fig:hidden_dimension}
\end{figure}

\Cref{fig:hidden_dimension} varies $H$ from one to eight. For all PanelGP
specifications, the rank correlation rises steeply up to two or three
factors and changes much less thereafter. The tangency Sharpe ratio keeps
rising beyond three factors; for RBF loadings and RBF factors, it increases
from $1.31$ at $H=3$ to $1.55$ at $H=5$. Additional
factors thus improve the covariance-based portfolio after the ranking has
stopped improving. The Gaussian log score improves more gradually through
$H=8$. Predictive $R^2$ behaves differently: for RBF/RBF it is
non-monotonic up to $H=5$ and changes little thereafter. The wide bands at
low dimensions partly reflect distinct solutions reached by different seeds,
and results become more stable as $H$ grows. We keep $H=5$, which was chosen
on the validation period.

\subsection{Annual re-estimation}
\label{sec:rolling}

A model estimated on data up to 1989 forecasts returns up to 35 years later.
We therefore also re-estimate the model with RBF loadings, RBF factors, and
$H=5$ once a year on the latest 360 input months. The first window runs from
January 1965 to December 1994 and produces the forecasts for the input months
of 1995; the last produces those for 2024. Each window has its own
initialization, context standardization, and return scale.

\begin{table}[!tb]
\centering
\caption{Fixed and annual rolling estimation with $H=5$ over the same 360
months. PanelGP uses RBF loadings and RBF factors, and its entries average
five seeds. Bold marks the best point estimate in each column; pairwise
tests are in \cref{tab:rolling_tests}.}
\label{tab:rolling}
\begingroup\small
\setlength{\tabcolsep}{4pt}\renewcommand{\arraystretch}{1.15}
\begin{tabular*}{\textwidth}{@{\extracolsep{\fill}}llrrrrrr@{}}
\toprule
 & & \multicolumn{3}{c}{Return forecasts} & \multicolumn{3}{c}{Portfolio performance} \\
\cmidrule(lr){3-5}\cmidrule(lr){6-8}
Model & Training & $R^2$ (\%) & Rank (\%) & Log score & Tangency & Long--short & GMV vol.\ (\%) \\
\midrule
PanelGP & Fixed & \textbf{0.753} & 4.52 & 1.092 & 1.55 & 0.98 & 14.3 \\
PanelGP & Rolling & 0.203 & \textbf{4.84} & \textbf{1.130} & \textbf{1.90} & \textbf{1.10} & \textbf{12.3} \\
\addlinespace[0.5ex]
IPCA & Fixed & 0.310 & 2.59 & 1.066 & 0.84 & 0.57 & 15.9 \\
IPCA & Rolling & 0.644 & 2.64 & 1.083 & 0.99 & 0.51 & 13.2 \\
\addlinespace[0.5ex]
IPCA+int. & Fixed & 0.285 & 3.11 & 1.061 & 1.17 & 0.65 & 16.3 \\
IPCA+int. & Rolling & 0.588 & 3.69 & 1.083 & 1.74 & 0.97 & 13.2 \\
\bottomrule
\end{tabular*}
\endgroup

\end{table}

Annual re-estimation raises the rank correlation of PanelGP from
4.52\% to 4.84\%, the log score from 1.092 to
1.130, and the tangency Sharpe ratio from 1.55 to
1.90, and it lowers GMV volatility from 14.3\% to
12.3\% (\cref{tab:rolling}), but only the reduction in GMV
volatility is significant (\cref{tab:rolling_tests}). Pooled predictive
$R^2$, by contrast, falls and ends up below both rolling IPCA benchmarks.
The fall is not significant (statistic 1.34), and it lies
in the level of the forecasts: the monthly average forecast contributes
0.744 percentage points under fixed training and
0.039 under annual re-estimation, while the deviations from it
contribute 0.009 and 0.164.
The strongest rolling benchmark is IPCA with an intercept. Rolling PanelGP
improves on it in three respects: a higher log score, a lower GMV
volatility, and a higher tangency Sharpe ratio, 1.90 against
1.74. The first two differences are significant;
the Sharpe ratios cannot be distinguished.

Every model in \cref{tab:rolling} has lower Sharpe ratios and rank
correlations in 2010--2024 than in 1995--2009 (\cref{tab:subperiods}). The
tangency Sharpe ratio of rolling PanelGP falls from
2.62 to 1.16 and is the highest
in both halves.

Sharpe ratios before costs ignore trading. Before rebalancing, the previous
weights drift with realized returns:
\begin{equation}
 \widetilde w_{i,t-1}
 =\frac{w_{i,t-1}(1+r_{f,t}+r_{i,t})}
 {1+r_{f,t}+\bm w_{t-1}^{\T}\bm r_t},\qquad
 \mathrm{TO}_t
 =\sum_{i\in\mathcal I_{t-1}\cup\mathcal I_t}
 |w_{i,t}-\widetilde w_{i,t-1}|,
 \label{eq:portfolio_turnover}
\end{equation}
where $r_{f,t}$ is the risk-free rate, $r_{i,t}$ the excess return, and
$w_{i,t}=0$ for $i\notin\mathcal I_t$. With a proportional cost $c$ per
dollar traded, the net return is
\begin{equation}
 r^{\mathrm{net}}_{p,t+1}
 =\bm w_t^{\T}\bm r_{t+1}-c\,\mathrm{TO}_t.
 \label{eq:net_portfolio_return}
\end{equation}
We set $c=10$ basis points, the order of institutional trading costs for
large stocks reported by \citet{frazzini_trading_2018}. A proportional cost
ignores stock-specific spreads, shorting fees, and price impact.

\begin{table}[!tb]
\centering
\caption{Trading costs under annual re-estimation with $H=5$. Turnover is the
average monthly $\mathrm{TO}_t$. Gross SR is the annualized Sharpe ratio
before costs, and Net SR the ratio after a cost of 10 basis points per dollar
traded. Break-even is the cost at which the average net excess return is
zero. PanelGP entries average five seeds.}
\label{tab:trading_costs}
\begingroup\small
\setlength{\tabcolsep}{4pt}\renewcommand{\arraystretch}{1.15}
\begin{tabular*}{\textwidth}{@{\extracolsep{\fill}}llrrrr@{}}
\toprule
Model & Portfolio & Turnover & Gross SR & Net SR, 10 bp & Break-even (bp) \\
\midrule
PanelGP & Tangency & 1.06 & 1.90 & 1.52 & 49.7 \\
PanelGP & Long--short & 1.38 & 1.10 & 0.91 & 57.3 \\
\addlinespace[0.5ex]
IPCA & Tangency & 0.64 & 0.99 & 0.85 & 70.1 \\
IPCA & Long--short & 0.98 & 0.51 & 0.36 & 34.4 \\
\addlinespace[0.5ex]
IPCA+int. & Tangency & 1.01 & 1.74 & 1.35 & 43.8 \\
IPCA+int. & Long--short & 1.38 & 0.97 & 0.72 & 38.5 \\
\bottomrule
\end{tabular*}
\endgroup

\end{table}

At 10 basis points, the tangency and long--short portfolios of rolling
PanelGP have net Sharpe ratios of 1.52 and
0.91, against 1.35 and
0.72 for rolling IPCA with an intercept
(\cref{tab:trading_costs}). Their average net excess returns stay positive up
to costs of 49.7 and 57.3 basis
points.

\subsection{Higher moments of the predictive distribution}
\label{sec:empirical_higher_moments}

This subsection and the next use the rolling forecasts. The bilinear
predictive distribution differs from the Gaussian distribution with the same
mean and variance only in its higher moments. For stock $i$ in month $t$, the
log-score gain from the higher moments is
\begin{equation}
 g_{it}=\log p_{\mathrm{bil}}(r_{i,t+1})
 -\log\Normal(r_{i,t+1}\given\hat\mu_{it},[\hat{\bm\Sigma}_t]_{ii}),
 \label{eq:bilinear_density_gain}
\end{equation}
where $p_{\mathrm{bil}}$ is the predictive density of the return, computed as
in \cref{app:marginal_density}. The gain averages about 0.022 per
observation and comes mostly from observations more than three standard
deviations from the mean, in particular in 2000--2002, 2008--2009, and 2020
(\cref{fig:density_gain}). Since the portfolios use only the first two
moments, this gain does not affect them.

\subsection{Predicted variance terms}
\label{sec:variance_mechanisms}

For stock $i$ in month $t$, we write $V^\phi_{it}$ and $V^\beta_t$ for the
terms in \cref{eq:innovation_variance_terms} and compare them with the
squared forecast errors $e_{it}^2=(r_{i,t+1}-\hat\mu_{it})^2$, which contain
both return risk and errors in the mean forecast. For a set $\mathcal G$ of
stock--months, the realized root mean squared (RMS) forecast error and the
predicted risk are
\begin{equation}
 \operatorname{RMS}(\mathcal G)
 =\left(\frac{1}{|\mathcal G|}\sum_{(i,t)\in\mathcal G}e_{it}^2\right)^{1/2},
 \qquad
 \widehat\sigma(\mathcal G)
 =\left(\frac{1}{|\mathcal G|}\sum_{(i,t)\in\mathcal G}
 [\hat{\bm\Sigma}_t]_{ii}\right)^{1/2}.
 \label{eq:risk_group_levels}
\end{equation}
Temporal quintiles weight months equally when averaging squared errors and
predicted variances, and cross-sectional quintiles pool stock--months.

The loading-innovation term $V^\beta_t$ is the same for all stocks in a
month. Its Spearman correlation over the 360 months with the monthly average
of $e_{it}^2$ is 58.3\% on average across seeds. The
factor-innovation term $V^\phi_{it}$ varies across stocks, and its Spearman
correlation with $e_{it}^2$ within a month, which removes market-wide
volatility, is 26.9\% on average over the 1,800 month--seed
pairs.

\begin{figure}[!tb]
\centering
\includegraphics[width=0.88\textwidth]{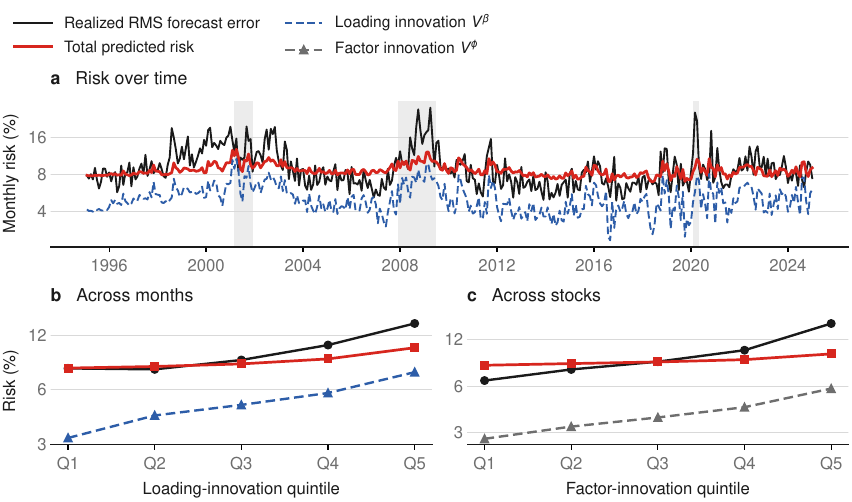}
\caption{Predicted variance terms under annual re-estimation. Top: realized
RMS forecast error, total predicted risk, and the loading-innovation term over
time; shading marks NBER peak-to-trough intervals. Bottom left: months sorted
by $V^\beta_t$. Bottom right: stocks sorted within each month by
$V^\phi_{it}$. Dates are return realization months. Variance terms are shown
as square roots, and vertical axes are logarithmic. Results pool five seeds.}
\label{fig:variance_evidence}
\end{figure}

\Cref{fig:variance_evidence} compares the ordering of risk with its level.
Across both months and stocks, realized RMS errors increase more steeply
than predicted risk, so the riskiest groups receive too little variance;
across stocks, the safest group also receives too much. Positive rank
correlations thus coexist with poorly calibrated risk levels. On average over
stock--months, $V^\beta_t$ is larger than $V^\phi_{it}$, and
$\tr(\bm D_\beta\bm D_\phi)$ is smaller than both. At a given date, $V^\beta_t$, $\tr(\bm D_\beta\bm D_\phi)$, and
$\sigma_r^2$ form a variance floor common to all stocks, which cannot adapt
to stock-specific idiosyncratic risk (\cref{sec:prediction}) and limits how
far predicted variances can spread across stocks.

\section{Conclusion}
\label{sec:conclusion}

The application shows that a single probabilistic factor model can deliver
competitive forecasts of both moments for a large panel of stocks.
Nonlinear loadings are the main source of improvement: whichever factor
kernel is used, they raise the rank correlation and the log score
significantly over affine loadings. Combined with context-dependent factors,
they give tangency portfolios with significantly higher Sharpe ratios and
minimum-variance portfolios with significantly lower volatility than IPCA.
Under annual re-estimation, the gains in log score and minimum-variance
volatility over IPCA with an intercept remain significant, and the tangency
portfolio keeps a Sharpe ratio of 1.52 after trading costs
of 10 basis points. These results come at moderate computational cost: a
five-factor fit takes about twelve minutes on one GPU.

The gains are concentrated in the ordering of stocks and in the covariance
matrix. In squared-error accuracy, IPCA does at least as well under annual
re-estimation, and the non-Gaussian shape of the predictive distribution
adds little to the log score. Every model has lower Sharpe ratios and rank
correlations in 2010--2024 than in 1995--2009, in line with the decline of
predictability from characteristics after the early 2000s.

The predicted variances order months and stocks by risk, but their level is
the main limitation: their dispersion is smaller than that of realized
errors because their idiosyncratic floor is common to all stocks at a date.
Letting $\bm D_\beta$ or $\sigma_r^2$ depend on the characteristics
would remove this common floor and keep the closed-form predictive law of
\cref{sec:prediction}, since predicted returns would still be quadratic
forms in a Gaussian state. Part of the compression may also come from the
reverse KL divergence of the variational approximation; other divergences
\citep{hernandezlobato_blackbox_2016,futami_robust_2018} could be compared on
this criterion.

\section*{Declarations}
\paragraph{Data and code availability.}
The stock characteristics \citep{jensen_is_2023} and the historical S\&P 500
membership (CRSP) are available to subscribers through Wharton Research Data
Services (WRDS) and are not redistributed; the macroeconomic data of
\citet{goyal_comprehensive_2024} are public. \ReplicationPackage{} contains
the code, a script that downloads and prepares all inputs, and the numerical
results behind every table and figure. Wharton Research Data Services (WRDS)
was used in preparing this paper. This service and the data
available thereon constitute valuable intellectual property and trade
secrets of WRDS and/or its third-party suppliers.

\paragraph{Funding.}
The authors received no specific funding for this work.

\paragraph{Disclosure statement.}
The authors report no competing interests.

\paragraph{Generative AI.}
Anthropic Claude (Claude Opus 5.5) assisted with software implementation,
numerical checks, plotting code, and revision of the manuscript. The authors
are responsible for the content of the manuscript and the accompanying code.

\clearpage
\appendix
\numberwithin{equation}{section}
\renewcommand{\thefigure}{S\arabic{figure}}
\renewcommand{\thetable}{S\arabic{table}}
\setcounter{figure}{0}
\setcounter{table}{0}
\begin{center}
\Large\bfseries Supplementary Material
\end{center}

\section{Variational inference}
\label{app:variational}

This appendix supports \cref{sec:variational_objective,sec:latent_prediction}.
\Cref{app:elbo_general} derives the variational identity,
\cref{app:elbo_gaussian} the terms of the objective for a single training
row, and \cref{app:predictive_conditioning} the predictive moments of the
latent state. The parameterization used in computation is in
\cref{app:computation}.

\subsection{Variational identity}
\label{app:elbo_general}

Fix $\bm\theta$. Let $q_{\bm\psi}$ be a distribution of the latent state
$\bm\chi$ in \cref{eq:latent_vector} that is absolutely continuous with
respect to the prior $p_{\bm\theta}(\bm\chi)$. By Bayes' theorem,
\begin{align}
 \KL\!\left(q_{\bm\psi}(\bm\chi)\|p_{\bm\theta}(\bm\chi\mid\bm r)\right)
 &=
 \E_q\!\left[
 \log\frac{q_{\bm\psi}(\bm\chi)\,p_{\bm\theta}(\bm r)}
 {p_{\bm\theta}(\bm r\mid\bm\chi)\,p_{\bm\theta}(\bm\chi)}
 \right]\notag\\
 &=
 \log p_{\bm\theta}(\bm r)
 -\E_q\!\left[\log p_{\bm\theta}(\bm r\mid\bm\chi)\right]
 +\KL\!\left(q_{\bm\psi}(\bm\chi)\|p_{\bm\theta}(\bm\chi)\right).
 \label{eq:posterior_kl_decomposition}
\end{align}
Rearranging gives
\begin{equation}
 \log p_{\bm\theta}(\bm r)
 =
 \mathcal L(\bm\theta,\bm\psi)
 +
 \KL\!\left(q_{\bm\psi}(\bm\chi)\|p_{\bm\theta}(\bm\chi\mid\bm r)\right),
 \label{eq:variational_gap}
\end{equation}
with $\mathcal L$ as in \cref{eq:main_variational_bound}. Since the returns
are independent across rows given $\bm B$ and $\widetilde{\bm F}$
(\cref{eq:likelihood}),
\begin{equation}
 \mathcal L(\bm\theta,\bm\psi)
 =
 \sum_{n=1}^{N}
 \E_q\!\left[
 \log p_{\bm\theta}
 \big(r_n\mid\bm\beta(\bm x_n),\widetilde{\bm\phi}_{t(n)}\big)
 \right]
 -
 \KL\!\left(q_{\bm\psi}(\bm\chi)\|p_{\bm\theta}(\bm\chi)\right).
 \label{eq:general_elbo_appendix}
\end{equation}
The divergence in \cref{eq:variational_gap} is nonnegative, so $\mathcal L$
is a lower bound on the log marginal likelihood $\log p_{\bm\theta}(\bm r)$.
For fixed $\bm\theta$, maximizing $\mathcal L$ over $\bm\psi$ minimizes this
divergence.

For the inducing-variable family in \cref{eq:q_joint}, the same identity
holds on the augmented state $(\bm B,\bm U,\widetilde{\bm F})$ and gives the
objective in \cref{eq:elbo}. Its gap is the divergence on the augmented
state, which is at least the divergence of the induced $q(\bm\chi)$ from
$p_{\bm\theta}(\bm\chi\mid\bm r)$, with equality when the rows of each
$\bm Z_h$ are training inputs, the inducing prior is nonsingular, and no
numerical stabilization is applied
\citep{titsias_variational_2009,matthews_sparse_2016}.

\subsection{Terms of the objective}
\label{app:elbo_gaussian}

Let $q_{\bm\psi}(\bm v)=\Normal(\bm m_v,\bm S_v)$ and
$p_{\bm\theta}(\bm v)=\Normal(\bm0,\widetilde{\bm K}_{vv})$, where
$\widetilde{\bm K}_{vv}$ is the prior covariance of the reduced state in
\cref{eq:reduced_latent_vector}, including the numerical stabilization of
\cref{app:stabilization}. The KL term of \cref{eq:elbo} is
\begin{equation}
 \KL\!\left(q_{\bm\psi}(\bm v)\|p_{\bm\theta}(\bm v)\right)
 =
 \frac12\left(
 \tr(\widetilde{\bm K}_{vv}^{-1}\bm S_v)
 +\bm m_v^{\T}\widetilde{\bm K}_{vv}^{-1}\bm m_v
 -d_v+\log\frac{|\widetilde{\bm K}_{vv}|}{|\bm S_v|}
 \right).
 \label{eq:gaussian_kl}
\end{equation}

Each likelihood term needs only the joint distribution under $q$ of the
smooth loading $\bm\beta(\bm x_n)$ and the realized factor
$\widetilde{\bm\phi}_t$ of row $n$, with $t=t(n)$. The factor is part of
$\bm v$, and the loading follows from the inducing values by Gaussian
conditioning. For component $h$ and an input matrix $\bm X_\circ$, define
\begin{align}
 \bm G_{\beta,h}(\bm X_\circ)
 &=
 \bm K_{\beta,h}(\bm X_\circ,\bm Z_h)
 \bm K_{\beta,h}(\bm Z_h,\bm Z_h)^{-1},\notag\\
 \bm R_{\beta,h}(\bm X_\circ)
 &=
 \bm K_{\beta,h}(\bm X_\circ,\bm X_\circ)
 -
 \bm G_{\beta,h}(\bm X_\circ)
 \bm K_{\beta,h}(\bm Z_h,\bm X_\circ).
 \label{eq:sparse_gp_maps}
\end{align}
Gaussian conditioning gives
\begin{equation}
 p_{\bm\theta}\!\left(\beta_h(\bm X_\circ)\mid\bm u_h\right)
 =
 \Normal\!\left(\bm G_{\beta,h}(\bm X_\circ)\bm u_h,
 \bm R_{\beta,h}(\bm X_\circ)\right),
 \label{eq:sparse_gp_conditional}
\end{equation}
where $\beta_h(\bm X_\circ)$ is the vector of values of $\beta_h$ at the rows
of $\bm X_\circ$; the same maps are used for prediction in
\cref{app:predictive_conditioning}. Partition the variational mean and
covariance of \cref{eq:q_joint} as
\begin{equation}
 \bm m_v=\begin{bmatrix}\bm m_u\\\bm m_\phi\end{bmatrix},\qquad
 \bm S_v=\begin{bmatrix}\bm S_{uu}&\bm S_{u\phi}\\
 \bm S_{\phi u}&\bm S_{\phi\phi}\end{bmatrix},
 \label{eq:q_blocks}
\end{equation}
and let $\bm m_{\phi,t}$, $\bm S_{u\phi,t}$, and $\bm S_{\phi\phi,t}$ be the
parts of $\bm m_\phi$, $\bm S_{u\phi}$, and $\bm S_{\phi\phi}$ that belong to
the $H$ coordinates of $\widetilde{\bm\phi}_t$. For the single input
$\bm x_n$, collect the components in
\begin{equation}
 \bm G_n=\operatorname{blockdiag}_{h=1}^{H}\{\bm G_{\beta,h}(\bm x_n^{\T})\}
 \in\R^{H\times MH},
 \qquad
 \bm R_n=\diag_{h=1}^{H}\{\bm R_{\beta,h}(\bm x_n^{\T})\}\in\R^{H\times H}.
 \label{eq:row_sparse_maps}
\end{equation}
Integrating \cref{eq:sparse_gp_conditional} against $q_{\bm\psi}(\bm v)$
gives
\begin{equation}
 q\!\left(\bm\beta(\bm x_n),\widetilde{\bm\phi}_{t}\right)
 =
 \Normal\!\left(
 \begin{bmatrix}\bm m_{\beta,n}\\\bm m_{\phi,t}\end{bmatrix},
 \begin{bmatrix}
 \bm S_{\beta\beta,n}&\bm S_{\beta\phi,n}\\
 \bm S_{\phi\beta,n}&\bm S_{\phi\phi,t}
 \end{bmatrix}
 \right),
 \qquad
 \begin{aligned}
 \bm m_{\beta,n}&=\bm G_n\bm m_u,\\
 \bm S_{\beta\beta,n}&=\bm R_n+\bm G_n\bm S_{uu}\bm G_n^{\T},\\
 \bm S_{\beta\phi,n}&=\bm G_n\bm S_{u\phi,t}.
 \end{aligned}
 \label{eq:row_joint}
\end{equation}
The off-diagonal blocks of $\bm S_{uu}$ and $\bm S_{u\phi,t}$ carry the
dependence across components and between loadings and factors. Each row
therefore needs only $\bm K_{\beta,h}(\bm x_n^{\T},\bm Z_h)$, the $M\times M$
matrices $\bm K_{\beta,h}(\bm Z_h,\bm Z_h)$, and the scalar prior variances
$\bm R_{\beta,h}(\bm x_n^{\T})$. Conditional on
$\widetilde{\bm\phi}_t=\bm\varphi$,
\begin{align}
 q\!\left(\bm\beta(\bm x_n)\mid\widetilde{\bm\phi}_t=\bm\varphi\right)
 &=\Normal\!\left(\bar{\bm\beta}_n(\bm\varphi),\bm V_n\right),
 \label{eq:conditional_beta}\\
 \bar{\bm\beta}_n(\bm\varphi)
 &=\bm m_{\beta,n}
 +
 \bm S_{\beta\phi,n}\bm S_{\phi\phi,t}^{-1}
 (\bm\varphi-\bm m_{\phi,t}),\notag\\
 \bm V_n
 &=\bm S_{\beta\beta,n}
 -\bm S_{\beta\phi,n}\bm S_{\phi\phi,t}^{-1}
 \bm S_{\phi\beta,n}.\notag
\end{align}
Let $\nu_{\bm\theta}(\bm\varphi)=\sigma_r^2+\bm\varphi^{\T}\bm D_\beta\bm\varphi$
be the likelihood variance in \cref{eq:marginal_likelihood}. It does not depend
on the loading, so the expectation over the loading is exact:
\begin{equation}
 \E_q\!\left[
 \log p_{\bm\theta}(r_n\mid\bm\beta(\bm x_n),\bm\varphi)
 \,\middle|\,
 \widetilde{\bm\phi}_t=\bm\varphi
 \right]
 =
 -\frac12\left[
 \log\!\left(2\pi \nu_{\bm\theta}(\bm\varphi)\right)
 +\frac{
 \left(r_n-\bar{\bm\beta}_n(\bm\varphi)^{\T}\bm\varphi\right)^2
 +\bm\varphi^{\T}\bm V_n\bm\varphi}
 {\nu_{\bm\theta}(\bm\varphi)}
 \right].
 \label{eq:expected_loglik}
\end{equation}
The remaining expectation over the $H$-dimensional
$q(\widetilde{\bm\phi}_t)$ is estimated by Monte Carlo with the
reparameterization $\widetilde{\bm\phi}_t=\bm m_{\phi,t}+\bm L_{\phi,t}\bm e$,
$\bm e\sim\Normal(\bm0,\bm I_H)$, where
$\bm L_{\phi,t}\bm L_{\phi,t}^{\T}=\bm S_{\phi\phi,t}$.

\subsection{Predictive moments}
\label{app:predictive_conditioning}

This subsection derives
\cref{eq:latent_gaussian_conditioning,eq:joint_latent_prediction}.
Conditioning on the whole reduced state $\bm v$ keeps all dependence across
components and between loadings and factors in $\bm S_v$.

For $h=1,\ldots,H$, let $\widetilde{\bm B}_{\star\cdot h}\in\R^J$ and
$\widetilde{\bm F}_{\star\cdot h}\in\R^W$ be the $h$th columns of
$\widetilde{\bm B}_\star$ and $\widetilde{\bm F}_\star$, so that
$\operatorname{vec}(\widetilde{\bm B}_\star)$ stacks the
$\widetilde{\bm B}_{\star\cdot h}$,
$\operatorname{vec}(\widetilde{\bm F}_\star)$ stacks the
$\widetilde{\bm F}_{\star\cdot h}$, and $\bm\chi_\star$ stacks
$\operatorname{vec}(\widetilde{\bm B}_\star)$ and
$\operatorname{vec}(\widetilde{\bm F}_\star)$ as in
\cref{eq:new_latent_vector}. Let $\bm X_\star$ be the $J\times P$
matrix of new characteristics and $\bm C_\star$ the $W\times Q$ matrix of
new contexts. Each new date has its own factor innovation, also when its
context equals a training context. The joint prior is
\begin{equation}
 \begin{bmatrix}\bm\chi_\star\\\bm v\end{bmatrix}
 \sim
 \Normal\!\left(
 \begin{bmatrix}\bm0\\\bm0\end{bmatrix},
 \begin{bmatrix}
  \widetilde{\bm K}_{\star\star}&\bm K_{\star v}\\
  \bm K_{\star v}^{\T}&\widetilde{\bm K}_{vv}
 \end{bmatrix}
 \right),
 \label{eq:joint_prediction_prior}
\end{equation}
with
\begin{align}
 \widetilde{\bm K}_{\star\star}
 &=
 \begin{bmatrix}
  \widetilde{\bm K}_{\beta\beta}^{\star\star}&0\\
  0&\widetilde{\bm K}_{\phi\phi}^{\star\star}
 \end{bmatrix},&
 \bm K_{\star v}
 &=
 \begin{bmatrix}
  \bm K_{\beta u}^{\star}&0\\
  0&\bm K_{\phi\phi}^{\star o}
 \end{bmatrix},&
 \widetilde{\bm K}_{vv}
 &=
 \begin{bmatrix}
  \bm K_{uu}&0\\
  0&\widetilde{\bm K}_{\phi\phi}^{oo}
 \end{bmatrix},
 \label{eq:joint_prediction_kernel_blocks}
\end{align}
and
\begin{equation}
\begin{aligned}
 \widetilde{\bm K}_{\beta\beta}^{\star\star}
 &=
 \operatorname{blockdiag}_{h}
 \left\{
 \bm K_{\beta,h}(\bm X_\star,\bm X_\star)
 +[\bm D_\beta]_{hh}\bm I_J
 \right\},\\
 \bm K_{\beta u}^{\star}
 &=
 \operatorname{blockdiag}_{h}
 \left\{
 \bm K_{\beta,h}(\bm X_\star,\bm Z_h)
 \right\},\\
 \bm K_{uu}
 &=
 \operatorname{blockdiag}_{h}
 \left\{
 \bm K_{\beta,h}(\bm Z_h,\bm Z_h)
 \right\},\\
 \widetilde{\bm K}_{\phi\phi}^{\star\star}
 &=
 \operatorname{blockdiag}_{h}
 \left\{
 \bm K_{\phi,h}(\bm C_\star,\bm C_\star)
 +[\bm D_\phi]_{hh}\bm I_W
 \right\},\\
 \bm K_{\phi\phi}^{\star o}
 &=
 \operatorname{blockdiag}_{h}
 \left\{
 \bm K_{\phi,h}(\bm C_\star,\bm C)
 \right\},\\
 \widetilde{\bm K}_{\phi\phi}^{oo}
 &=
 \operatorname{blockdiag}_{h}
 \left\{
 \bm K_{\phi,h}(\bm C,\bm C)
 +[\bm D_\phi]_{hh}\bm I_T
 \right\},
\end{aligned}
\label{eq:joint_prediction_block_details}
\end{equation}
where $\operatorname{blockdiag}_h$ runs over $h=1,\ldots,H$.
The new innovations enter $\widetilde{\bm K}_{\star\star}$ but not
$\bm K_{\star v}$, because they are independent of the training state. To
reconstruct a realized training factor instead, its coordinates are taken
directly from $\bm v$.

Conditioning the prior in \cref{eq:joint_prediction_prior} on $\bm v$ gives
\cref{eq:latent_gaussian_conditioning}, and integrating over
$q_{\bm\psi}(\bm v)$ gives \cref{eq:joint_latent_prediction}. Since
$\bm K_{\star v}$ and $\widetilde{\bm K}_{vv}$ are block diagonal, so are
$\bm G_\star=\operatorname{blockdiag}(\bm G_\beta^\star,\bm G_\phi^\star)$ and
$\bm R_\star=\operatorname{blockdiag}(\bm R_\beta^\star,\bm R_\phi^\star)$.
Here $\bm R_\beta^\star=\operatorname{blockdiag}_h\{\bm R_{\beta,h}(\bm X_\star)\}
+\bm D_\beta\otimes\bm I_J$, with $\bm R_{\beta,h}$ from
\cref{eq:sparse_gp_maps}. With the partition in \cref{eq:q_blocks},
\begin{align}
 \bm m_\beta^\star
 &=\bm G_\beta^\star\bm m_u,&
 \bm m_\phi^\star
 &=\bm G_\phi^\star\bm m_\phi,\notag\\
 \widetilde{\bm S}_{\beta\beta}^\star
 &=\bm R_\beta^\star
 +\bm G_\beta^\star \bm S_{uu}(\bm G_\beta^\star)^{\T},&
 \widetilde{\bm S}_{\phi\phi}^\star
 &=\bm R_\phi^\star
 +\bm G_\phi^\star \bm S_{\phi\phi}(\bm G_\phi^\star)^{\T},\notag\\
 \widetilde{\bm S}_{\beta\phi}^\star
 &=\bm G_\beta^\star \bm S_{u\phi}(\bm G_\phi^\star)^{\T}.&
 \label{eq:predictive_block_moments_appendix}
\end{align}
The prior blocks are diagonal across components, but the off-diagonal blocks
of $\bm S_{uu}$, $\bm S_{\phi\phi}$, and $\bm S_{u\phi}$ create predictive
dependence across components. Selecting rows and dates gives the vectors
$\bm m_{\beta,j}$ and $\bm m_{\phi,t}$ and the blocks
$[\widetilde{\bm S}_\star]_{\beta_j\beta_{j'}}$,
$[\widetilde{\bm S}_\star]_{\phi_t\phi_{t'}}$, and
$[\widetilde{\bm S}_\star]_{\beta_j\phi_t}$ used in
\cref{eq:predictive_mean} and \cref{eq:predictive_covariance}.

\section{Predictive return distributions}
\label{app:quadratic}

This appendix proves the results of \cref{sec:prediction}.
\Cref{app:new_panel_law} derives the joint CGF of a new return panel and
the predictive mean and covariance that follow from it,
\cref{app:scalar_functionals} the cumulants of portfolio returns and the
tail behavior they imply, and \cref{app:marginal_density} the numerical
evaluation of the predictive density of a single return.

\subsection{Joint CGF of a new return panel}
\label{app:new_panel_law}
\label{app:quadratic_cgf}

Consider the $J$ new rows and $W$ new dates of \cref{sec:latent_prediction},
with $\bm\chi_\star\sim\Normal(\bm m_\star,\widetilde{\bm S}_\star)$ and
$d_\star=\dim(\bm\chi_\star)=(J+W)H$. Let $\bm\Pi$ be the $2HJ\times d_\star$
matrix with entries in $\{0,1\}$ that arranges the state by row:
\begin{equation}
 \bm\Pi\bm\chi_\star
 =
 \begin{bmatrix}
  \widetilde{\bm\beta}_{\star1}^{\T}&
  \widetilde{\bm\phi}_{\star t(1)}^{\T}&
  \cdots&
  \widetilde{\bm\beta}_{\star J}^{\T}&
  \widetilde{\bm\phi}_{\star t(J)}^{\T}
 \end{bmatrix}^{\T}.
 \label{eq:panel_permutation}
\end{equation}
When several rows share a date, $\bm\Pi$ repeats the factor coordinates of
that date; it is a permutation matrix only when $W=J$. Define
\begin{equation}
\begin{gathered}
 \bm\Omega_H=\frac12
 \begin{bmatrix}
 0&\bm I_H\\
 \bm I_H&0
 \end{bmatrix},
 \qquad
 \bm A_j=\bm\Pi^{\T}\left(\bm e_j\bm e_j^{\T}\otimes\bm\Omega_H\right)\bm\Pi,\\
 \bm A(\bm s)=\sum_{j=1}^Js_j\bm A_j
 =\bm\Pi^{\T}\left(\diag(\bm s)\otimes\bm\Omega_H\right)\bm\Pi,
\end{gathered}
 \label{eq:panel_quadratic_matrix}
\end{equation}
where $\bm e_j$ is the $j$th unit vector of $\R^J$. The matrix $\bm A(\bm s)$
is symmetric and linear in $\bm s$, and
\begin{equation}
 \bm s^{\T}\bm r_\star
 =
 \sum_{j=1}^J
 s_j\widetilde{\bm\beta}_{\star j}^{\T}
 \widetilde{\bm\phi}_{\star t(j)}
 +\sigma_r\bm s^{\T}\bm\epsilon_\star
 =
 \bm\chi_\star^{\T}\bm A(\bm s)\bm\chi_\star
 +\sigma_r\bm s^{\T}\bm\epsilon_\star,
 \qquad
 \bm\epsilon_\star\sim\Normal(\bm 0,\bm I_J).
 \label{eq:new_panel_quadratic}
\end{equation}
For fixed $\bm s$, $\bm s^{\T}\bm r_\star$ is a single Gaussian quadratic form
plus independent Gaussian noise, so the classical result
\citep{mathai_quadratic_1992} applies with the matrix $\bm A(\bm s)$. We give
the derivation for completeness.

The matrix $\widetilde{\bm S}_\star$ is positive definite, because $\bm R_\star$
is at least $\operatorname{blockdiag}(\bm D_\beta\otimes\bm I_J,
\bm D_\phi\otimes\bm I_W)$ in the positive semidefinite order. By independence
of $\bm\chi_\star$ and $\bm\epsilon_\star$,
\begin{multline}
 \E\left[\exp(\bm s^{\T}\bm r_\star)\right]
 =
 \exp\!\left(\frac12\sigma_r^2\bm s^{\T}\bm s\right)
 \frac{1}{(2\pi)^{d_\star/2}|\widetilde{\bm S}_\star|^{1/2}}\\
 \times\int_{\R^{d_\star}}
 \exp\!\left[
 -\frac12(\bm\chi-\bm m_\star)^{\T}
 \widetilde{\bm S}_\star^{-1}(\bm\chi-\bm m_\star)
 +\bm\chi^{\T}\bm A(\bm s)\bm\chi
 \right]d\bm\chi .
 \label{eq:new_panel_mgf_integral}
\end{multline}
The integral is finite if and only if
\begin{equation}
 \widetilde{\bm S}_\star^{-1}-2\bm A(\bm s)\succ0 .
 \label{eq:cgf_domain_matrix}
\end{equation}
This set of $\bm s$ is open and convex, because $\bm A(\bm s)$ is linear in
$\bm s$, and it contains $\bm 0$. On it, completing the square and simplifying
the determinant gives
\begin{equation}
 \E\left[\exp(\bm s^{\T}\bm r_\star)\right]
 =
 |\bm I-2\widetilde{\bm S}_\star \bm A(\bm s)|^{-1/2}
 \exp\!\left[
 \bm m_\star^{\T}\bm A(\bm s)
 \left(\bm I-2\widetilde{\bm S}_\star \bm A(\bm s)\right)^{-1}
 \bm m_\star
 +\frac12\sigma_r^2\bm s^{\T}\bm s
 \right].
 \label{eq:quadratic_mgf}
\end{equation}
Taking logarithms gives \cref{eq:cgf} and proves \cref{prop:joint_cgf}:
\begin{equation}
 \mathcal K(\bm s)
 =
 -\frac12\log|\bm I-2\widetilde{\bm S}_\star \bm A(\bm s)|
 +
 \bm m_\star^{\T}\bm A(\bm s)
 \left(\bm I-2\widetilde{\bm S}_\star \bm A(\bm s)\right)^{-1}
 \bm m_\star
 +\frac12\sigma_r^2\bm s^{\T}\bm s.
 \label{eq:joint_cgf_appendix}
\end{equation}
Since $\mathcal K$ is finite on a neighborhood of zero, it is analytic there,
and its mixed derivatives at zero give the joint cumulants of the new
returns:
\begin{equation}
 \operatorname{cum}(r_{\star j_1},\ldots,r_{\star j_k})
 =
 \left.
 \frac{\partial^k\mathcal K(\bm s)}
 {\partial s_{j_1}\cdots\partial s_{j_k}}
 \right|_{\bm s=\bm 0}.
 \label{eq:mixed_cumulants}
\end{equation}
The first derivatives give the predictive mean in \cref{eq:predictive_mean},
$\hat\mu_{\star j}
=\E_q(\widetilde{\bm\beta}_{\star j}^{\T}\widetilde{\bm\phi}_{\star t(j)})
=\bm m_{\beta,j}^{\T}\bm m_{\phi,t(j)}
+\tr([\widetilde{\bm S}_\star]_{\beta_j\phi_{t(j)}})$. The second
derivatives, or equivalently Isserlis' theorem applied to the products in
\cref{eq:quadratic_form}, give the predictive covariance. With the block
notation of \cref{sec:prediction}, for rows $j$ and $j'$ with new dates
$t=t(j)$ and $t'=t(j')$,
\begin{align}
 \hat\Sigma_{\star,jj'}
 &=
 \bm m_{\beta,j}^{\T}[\widetilde{\bm S}_\star]_{\phi_t\phi_{t'}}\bm m_{\beta,j'}
 +\bm m_{\phi,t}^{\T}[\widetilde{\bm S}_\star]_{\beta_j\beta_{j'}}\bm m_{\phi,t'}
 +\bm m_{\beta,j}^{\T}[\widetilde{\bm S}_\star]_{\phi_t\beta_{j'}}\bm m_{\phi,t'}
 +\bm m_{\phi,t}^{\T}[\widetilde{\bm S}_\star]_{\beta_j\phi_{t'}}\bm m_{\beta,j'}
 \notag\\
 &\quad
 +\tr\!\left(
 [\widetilde{\bm S}_\star]_{\beta_j\beta_{j'}}^{\T}
 [\widetilde{\bm S}_\star]_{\phi_t\phi_{t'}}
 \right)
 +\tr\!\left(
 [\widetilde{\bm S}_\star]_{\beta_j\phi_{t'}}
 [\widetilde{\bm S}_\star]_{\phi_t\beta_{j'}}^{\T}
 \right)
 +\sigma_r^2 \delta_{jj'}.
 \label{eq:predictive_covariance}
\end{align}
In particular, $\Var_q(r_{\star j})=\hat\Sigma_{\star,jj}$.

\subsection{Cumulants of portfolio returns}
\label{app:scalar_functionals}

For weights $\bm w\in\R^J$, the return
$r_{\bm w}=\bm w^{\T}\bm r_\star
=\bm\chi_\star^{\T}\bm A(\bm w)\bm\chi_\star+\sigma_r\bm w^{\T}\bm\epsilon_\star$
has the scalar CGF $\mathcal K_{\bm w}(s)=\mathcal K(s\bm w)$:
\begin{equation}
 \mathcal K_{\bm w}(s)
 =
 -\frac12\log|\bm I-2s\widetilde{\bm S}_\star \bm A(\bm w)|
 +
 s\,\bm m_\star^{\T}\bm A(\bm w)
 (\bm I-2s\widetilde{\bm S}_\star \bm A(\bm w))^{-1}\bm m_\star
 +\frac12s^2\sigma_r^2\|\bm w\|_2^2.
 \label{eq:scalar_cgf}
\end{equation}
For $|s|<1/(2\varrho)$, where $\varrho$ is the spectral radius of
$\widetilde{\bm S}_\star\bm A(\bm w)$, the expansions
\begin{equation}
 -\log|\bm I-\bm Y|
 =\sum_{k=1}^{\infty}\frac{\tr(\bm Y^k)}{k},
 \qquad
 (\bm I-\bm Y)^{-1}=\sum_{k=0}^{\infty}\bm Y^k,
 \qquad \bm Y=2s\widetilde{\bm S}_\star\bm A(\bm w),
 \label{eq:cgf_series}
\end{equation}
converge. Matching the coefficients of $s^k/k!$ gives, for every $k\ge1$,
\begin{equation}
 \kappa_k(r_{\bm w})
 =
 2^{k-1}(k-1)!\left[
 \tr\!\left((\bm A(\bm w)\widetilde{\bm S}_\star)^k\right)
 +k\,\bm m_\star^{\T}
 (\bm A(\bm w)\widetilde{\bm S}_\star)^{k-1}
 \bm A(\bm w)\bm m_\star
 \right]
 +
 \delta_{k2}\sigma_r^2\|\bm w\|_2^2,
 \label{eq:all_cumulants_appendix}
\end{equation}
which proves \cref{prop:all_cumulants}.

The scalar CGF also determines the tails of the portfolio return. Let
$\lambda_1,\ldots,\lambda_{d_\star}$ be the eigenvalues of
$\widetilde{\bm S}_\star^{1/2}\bm A(\bm w)\widetilde{\bm S}_\star^{1/2}$. The
scalar CGF in \cref{eq:scalar_cgf} is finite if and only if $2s\lambda_i<1$ for
all $i$, which holds on an open interval around zero. For any $s>0$ in this
interval, Markov's inequality gives
$P(r_{\bm w}>x)\le\exp(\mathcal K_{\bm w}(s)-sx)$, and the lower tail is
bounded in the same way with $s<0$. Both tails therefore decay at least
exponentially. A power-law tail has no finite exponential moment, so the
predictive distribution can be skewed and have excess kurtosis but cannot
have power-law tails. Heavier tails would require a different noise
distribution, for example a Student-$t$ observation noise. The CGF would then
no longer exist, and the closed form of \cref{prop:joint_cgf} would be lost.

\subsection{Predictive density of a single return}
\label{app:marginal_density}

The log-score gains in \cref{fig:density_gain} use the predictive density of a
single return. Its evaluation reduces to one-dimensional Fourier inversion
of the CGF, regardless of the number of factors. Consider row $j$ at new
date $t=t(j)$. By \cref{eq:joint_latent_prediction},
\begin{equation}
 \begin{bmatrix}\widetilde{\bm\beta}_{\star j}\\ \widetilde{\bm\phi}_{\star t}\end{bmatrix}
 \sim\Normal\!\left(
 \begin{bmatrix}\bm m_{\beta,j}\\ \bm m_{\phi,t}\end{bmatrix},
 \begin{bmatrix}
  [\widetilde{\bm S}_\star]_{\beta_j\beta_j}&[\widetilde{\bm S}_\star]_{\beta_j\phi_t}\\
  [\widetilde{\bm S}_\star]_{\phi_t\beta_j}&[\widetilde{\bm S}_\star]_{\phi_t\phi_t}
 \end{bmatrix}\right).
 \label{eq:predictive_row_joint}
\end{equation}
This $2H$-dimensional Gaussian state includes both innovations and the full
loading--factor covariance. Its quadratic form with matrix $\bm\Omega_H$
in \cref{eq:panel_quadratic_matrix}, plus independent observation noise,
gives $r_{\star j}$. Let $K_j(s)=\mathcal K(s\bm e_j)$ be its scalar CGF
from \cref{eq:joint_cgf_appendix}.

For a real $s$ inside the CGF domain, the exponentially tilted density
$p_s(x)=\exp\{sx-K_j(s)\}p_{\mathrm{bil}}(x)$ integrates to one. Its
characteristic function is $\exp\{K_j(s+iu)-K_j(s)\}$.
Fourier inversion therefore gives
\begin{equation}
 p_{\mathrm{bil}}(r)=\frac{e^{K_j(s)-sr}}{\pi}
 \int_0^\infty\operatorname{Re}\!\left[
 e^{K_j(s+iu)-K_j(s)-iur}\right]du.
 \label{eq:marginal_return_density}
\end{equation}
The integrand is integrable because the observation variance is positive.
The identity holds for every admissible $s$, and the choice of $s$ affects
only the numerical scaling.

For each observed return, we solve $K_j'(s)=r$ inside the CGF domain and set
$d=\sqrt{K_j''(s)}$. The change of variable $u=w/d$ gives the log density
\begin{equation}
 \log p_{\mathrm{bil}}(r)=K_j(s)-sr-\log(\pi d)
 +\log\!\int_0^\infty\operatorname{Re}\!\left[
 e^{K_j(s+iw/d)-K_j(s)-iwr/d}\right]dw.
 \label{eq:marginal_density_quadrature}
\end{equation}
The root centers the tilted distribution at $r$, and the rescaling gives it
unit variance, which stabilizes evaluation in the tails. The integral is
evaluated adaptively with SciPy's \texttt{quad\_vec}, with absolute and
relative tolerances $10^{-9}$. The density is thus obtained by numerical
inversion of the fitted law, without a Gaussian or saddlepoint
approximation. Computation uses scaled returns; subtracting $\log s_r$
converts the log density to original return
units, consistently with \cref{app:target_scale}.

\section{Model specification and computation}
\label{app:specification}

This appendix gives the settings needed to reproduce the simulation and the
application. \Cref{app:target_scale} describes the scaling of inputs and
returns, \cref{app:kernel_specification} the parameterization of the kernels
and innovation variances, \cref{app:computation} the computational form of
the variational family, and \cref{app:settings} the initialization and
optimization.

\subsection{Input and return scaling}
\label{app:target_scale}

Characteristics are ranked within each month and mapped to $[-0.5,0.5]$, and
missing values are set to zero. The ranks use only the current month, so the
characteristics of a new month lie in the same range as those of the training
sample. The context is standardized to zero mean and unit variance with
training-sample moments, and a new context can lie outside the training
range.

Returns are scaled without centering,
\begin{equation}
 r_n^\circ=r_n/s_r,\qquad s_r^2=\frac{1}{N_{\mathrm{train}}}
 \sum_{n\in\mathrm{train}}r_n^2,
 \label{eq:asset_target_normalization}
\end{equation}
where $N_{\mathrm{train}}$ is the number of training rows. The scaling does not
restrict the expected return. Before evaluation, predictive means are
multiplied by $s_r$, covariances by $s_r^2$, and cumulants of order $k$ by
$s_r^k$. In the alignment of the simulation, the factor $s_r$ is absorbed into
the loadings, so $\bm D_\beta$ is multiplied by $s_r^2$.

In the simulation, characteristics and context lie in $[0,1]$ and are used
without ranking or standardization.

\subsection{Kernels and innovation variances}
\label{app:kernel_specification}

For a vector of lengthscales $\bm\ell$, the RBF kernel is
\begin{equation}
 \bar k_{\mathrm{RBF}}(\bm x,\bm x';\bm\ell)=
 \exp\!\left[-\frac12\sum_{p}
 \frac{(x_p-x'_p)^2}{\ell_p^2}\right].
 \label{eq:focused_loading_kernels}
\end{equation}
The affine kernel is $1+\bm x^{\T}\bm x'$, and the constant kernel is $1$.

\paragraph{Loadings.} Each specification uses the same kernel family for all
components, with component-specific parameters:
\begin{equation}
 k_{\beta,h}(\bm x,\bm x')=\eta_h\bar k_{\beta,h}(\bm x,\bm x'),
 \qquad
 \bar k_{\beta,h}(\bm x,\bm x')\in\left\{1+\bm x^{\T}\bm x',\;
 \bar k_{\mathrm{RBF}}(\bm x,\bm x';\bm\ell_{\beta,h})\right\},
 \label{eq:loading_covariance}
\end{equation}
with lengthscales $\bm\ell_{\beta,h}\in\R^P$ for the RBF kernel. The output
scale $\eta_h$ and the loading-innovation variance
$\tau_{\beta,h}=[\bm D_\beta]_{hh}$ are softplus transforms of unconstrained
parameters, which keeps them positive during training.

\paragraph{Factors.} The factor kernel in \cref{eq:factor_kernel} has
lengthscales $\bm\ell_{\phi,h}\in\R^Q$ and gives the realized-factor
covariance
\begin{equation}
 \widetilde k_{\phi,h}(\bm c_t,\bm c_{t'})
 =a_h \bar k_{\mathrm{RBF}}(\bm c_t,\bm c_{t'};\bm\ell_{\phi,h})+b_h+\rho_h\delta_{tt'},
 \qquad a_h+b_h+\rho_h=1,
 \label{eq:implemented_factor_covariance}
\end{equation}
where $\rho_h=[\bm D_\phi]_{hh}$. In the application, a three-way softmax
parameterizes the shares $(a_h,b_h,\rho_h)$ of the RBF factor kernel. The
constant factor kernel sets $a_h=0$ and parameterizes $\rho_h$ by a sigmoid,
with $b_h=1-\rho_h$. Each realized factor has unit prior variance, and the
scale of returns is carried by the loading output scale $\eta_h$.

The RBF factor kernel has a constant term, and the RBF loading kernel does
not. The reason is extrapolation. Characteristic ranks are bounded, but a
standardized context can lie far from all training contexts, where the RBF
term vanishes. Without the constant term, the factor forecast would then
tend to zero. With it, the forecast tends to
$b_h\bm 1^{\T}\widetilde{\bm K}_{\phi,h}^{-1}\bm m_{\phi}^{(h)}$, a shrunken
generalized least-squares average of the fitted training factors, where
$\widetilde{\bm K}_{\phi,h}$ is the prior covariance in
\cref{eq:factor_gp_prior} and $\bm m_{\phi}^{(h)}$ the variational mean of
$\widetilde{\bm F}_{\cdot h}$.

\paragraph{Simulation.} The simulation uses RBF kernels for loadings and
factors, and its factor kernel keeps the constant term. To match the
isotropic design, the loading-innovation variance and the factor-innovation
variance are each shared by the two components. The shared factor-innovation
share is parameterized by a sigmoid and the two remaining smooth shares by a
two-way softmax, so the smooth kernels remain component-specific.

\subsection{Computation}
\label{app:computation}

\paragraph{Variational state.}
The inducing inputs are initialized at distinct training rows and then
optimized jointly with the model parameters. There are 160 per component in
the application and 64 in the simulation. The factor state contains all
training dates, so the reduced state $\bm v$ has dimension $d_v=(M+T)H$; with
$M=160$, $T=360$, and $H=5$, $d_v=2{,}600$. The affine kernel spans only
138 functions (137 characteristics and a
constant), fewer than the number of inducing inputs, so the inducing
variables do not restrict the affine loadings.

\paragraph{Whitened parameterization.}
\label{app:whitening}
The computations use a whitened state. Let
$\widetilde{\bm K}_{vv}=\bm L_v\bm L_v^{\T}$ and $\bm v=\bm L_v\bm\omega$, so
that $\bm\omega\sim\Normal(\bm 0,\bm I)$ under the prior. The variational
distribution is $q_{\bm\psi}(\bm\omega)=\Normal(\bm m_\omega,\bm S_\omega)$,
which corresponds to $\bm m_v=\bm L_v\bm m_\omega$ and
$\bm S_v=\bm L_v\bm S_\omega\bm L_v^{\T}$, and $\bm S_\omega$ has a dense
trainable Cholesky factor. The divergence in \cref{eq:gaussian_kl} becomes
\begin{equation}
 \KL\!\left(q_{\bm\psi}(\bm\omega)\|\Normal(\bm 0,\bm I)\right)
 =
 \frac12\left(
 \tr(\bm S_\omega)+\bm m_\omega^{\T}\bm m_\omega-d_v-\log|\bm S_\omega|
 \right).
 \label{eq:whitened_kl}
\end{equation}
The reparameterization leaves the model, the objective, and the predictive
distribution unchanged. It improves numerical conditioning, especially for
the affine kernel, whose inducing covariance is nearly singular.

\paragraph{Numerical stabilization.}
\label{app:stabilization}
Prior covariance inverses use Cholesky factorizations with a diagonal term
that starts at $10^{-7}$ and increases tenfold until the factorization
succeeds. This matters for the affine loading kernel, whose $160\times160$
inducing covariance has rank at most 138. The stabilized
$\bm K_{uu}$ then replaces the prior covariance in $\bm G_\star$,
$\bm R_\star$, and $p_{\bm\theta}(\bm U)$. This is equivalent to observing
$\bm U$ with a small independent noise, so the objective in \cref{eq:elbo}
remains a lower bound.

\subsection{Initialization and optimization}
\label{app:settings}

\paragraph{Variances.}
The initial observation variance is $\sigma_r^2=0.1$ in scaled units. The
loading output scales start equal across components, with
$\tau_{\beta,h}=\gamma_\beta\eta_h$ and $\gamma_\beta=0.1$, and are set so
that the average prior variance of the scaled return is one:
\begin{equation}
 \sigma_r^2+\sum_{h=1}^H\left(\eta_h d+\tau_{\beta,h}\right)=1,
 \qquad\text{so}\qquad
 \eta_h=\frac{1-\sigma_r^2}{H(d+\gamma_\beta)}=\frac{0.9}{H(d+0.1)}.
 \label{eq:loading_scale_budget}
\end{equation}
Here $d$ is the training average of the unscaled loading-kernel diagonal:
$d=1$ for the RBF kernel and
$d=1+\operatorname{mean}_{\rm train}\|\bm x\|^2$ for the affine kernel. With
unit factor prior variances, both loading kernels thus start from the same
return-variance budget. For the RBF factor kernel, the initial shares are
$\rho_h=0.1$, $b_h=0.1$, and $a_h=0.8$. All these parameters are estimated
after initialization.

\paragraph{Lengthscales.}
All lengthscales start from the same rule. For two independent inputs
$\bm x,\bm x'$ with coordinate variances $\varsigma_p^2$,
\begin{equation}
 \E\|\bm x-\bm x'\|^2=2\sum_p \varsigma_p^2 .
 \label{eq:bandwidth_initialization}
\end{equation}
Setting every initial lengthscale to $\ell_0=(\sum_p\varsigma_p^2)^{1/2}$ makes
the RBF kernel equal to $e^{-1}$ at the expected squared distance.
Characteristic ranks are approximately uniform on $[-0.5,0.5]$, with variance
$1/12$, so $\ell_0=\sqrt{P/12}\approx3.38$ for
$P=137$. Standardized context variables have unit variance, so
$\ell_0=\sqrt Q=2.45$ for $Q=6$. In the simulation, inputs lie in $[0,1]$, and
the same variance $1/12$ gives $\sqrt{2/12}\approx0.408$ for the two
characteristics and $\sqrt{1/12}\approx0.289$ for the context.

\paragraph{Optimization.}
Both experiments initialize the whitened mean with independent Gaussian
perturbations of standard deviation $0.001$ and the covariance at the
identity. Adam uses an initial learning rate of $0.01$ for all variational
parameters and $0.005$ for the model parameters, including the inducing
inputs, and gradients are clipped to unit Euclidean norm. Training uses
minibatches of 2,048 rows and 64 factor samples per likelihood evaluation,
formed from 32 independent Gaussian draws and their negatives; each factor
draw is shared by all rows of its date. The application uses 500 epochs and
the simulation 5{,}000. Learning rates halve at 30\%, 50\%, 60\%, 70\%,
80\%, and 90\% of training, and the final iterate is used. A five-factor
empirical fit takes about twelve minutes on an NVIDIA H100 GPU.

The training objective is nonconvex, and the low-dimensional empirical fits
sometimes reach distinct solutions for different seeds, which widens the
bands in \cref{fig:hidden_dimension}. All five seeds are retained, so the
reported dispersion describes optimization sensitivity on the same return
panel.

\section{Additional results}
\label{app:additional_results}
\FloatBarrier

This appendix reports results that support the simulation and the
application, in the order in which the main text refers to them: loading
recovery in the simulation (\cref{app:loading_recovery}), validation results
by number of factors (\cref{app:validation}), pairwise tests between models
(\cref{app:statistical_comparisons}), results by subperiod
(\cref{app:subperiods}), and the log-score gain of the bilinear predictive
density (\cref{app:density_gain}).

\subsection{Loading recovery in the simulation}
\label{app:loading_recovery}

\begin{figure}[!tb]
\centering
\includegraphics[width=0.9\textwidth]{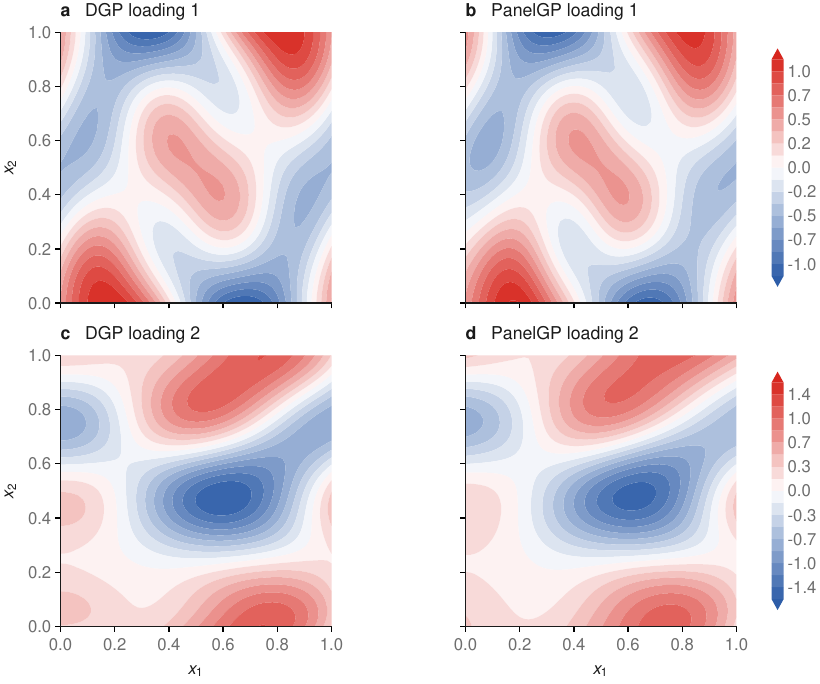}
\caption{Loading recovery in simulation replication~4{}.
Rows correspond to the two latent components; columns show the true surface
and the aligned fitted mean, both evaluated on the same characteristic
grid. Colors use a common scale within each row.}
\label{fig:synthetic_loadings}
\end{figure}

The loading functions depend on two characteristics, so they are surfaces,
whereas the factor responses in \cref{fig:synthetic_factors} are curves.
\Cref{fig:synthetic_loadings} compares the true
surfaces with the fitted smooth loading means after the factor alignment
described in \cref{sec:synthetic}, for the same replication as in
\cref{fig:synthetic_factors}. The fitted surfaces reproduce the locations and
signs of the peaks and troughs, and the common color scale within each row
shows that the magnitudes match as well. The figure shows the smooth
loadings $\bm\beta(\bm x)$; the realized loadings
$\widetilde{\bm\beta}_{it}$ also contain innovations.

\subsection{Validation period and the number of factors}
\label{app:validation}

\begin{figure}[!tb]
\centering
\includegraphics[width=\textwidth]{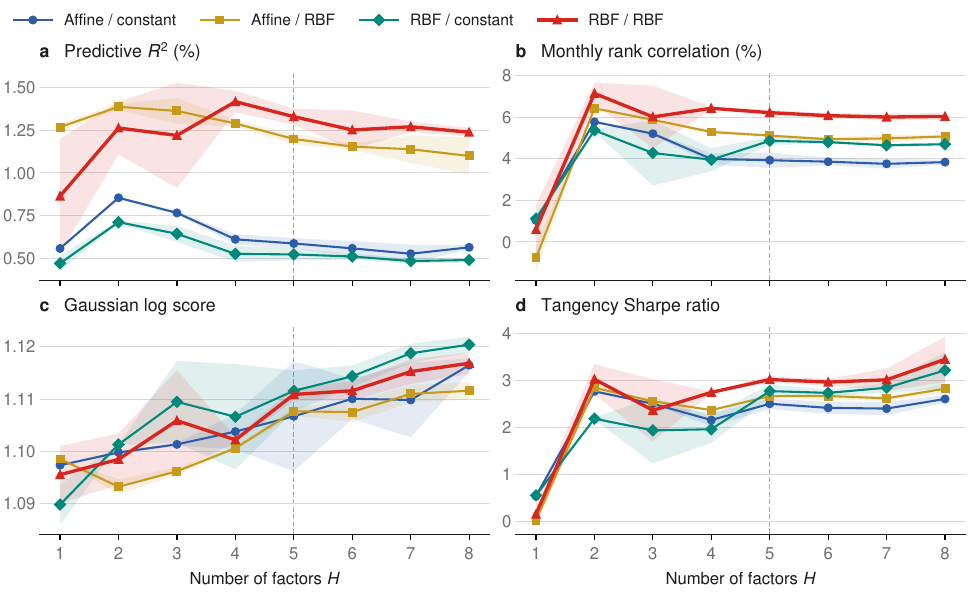}
\caption{Validation performance by number of factors $H$. All models are
estimated on 1960--1989 and evaluated on 1990--1994. Colors and markers match
\cref{fig:hidden_dimension}. Curves show means across five optimization
seeds, and bands extend one sample standard deviation on either side. The
dashed vertical line marks $H=5$. Sharpe ratios are before trading costs.}
\label{fig:validation}
\end{figure}

\Cref{fig:validation} shows the four PanelGP kernel pairs on the validation
period, 1990--1994, as the number of factors increases. The criteria favor
different dimensions. The Gaussian log score keeps improving beyond the
dimensions at which the mean-forecast criteria level off, while each
additional factor enlarges the dense variational covariance. We use $H=5$ as
a compromise between validation performance and model complexity.

\subsection{Pairwise model comparisons}
\label{app:statistical_comparisons}

Within each of \cref{tab:focused_kernels,tab:rolling}, we compare every pair
of rows on every criterion. All comparisons use the same calendar months and
Bartlett long-run variance estimates with twelve lags
\citep{newey_west_1987}.

For predictive $R^2$, the comparison is based on monthly total squared
errors; since the full-sample return sum of squares is the same for both
models, the average difference in squared errors maps into the difference in
pooled $R^2$. For rank correlation and the Gaussian log score, we compare
monthly scores. PanelGP quantities are averaged over the five optimization
seeds before the monthly difference is formed.

Sharpe ratios and volatilities are nonlinear functions of returns. We
compare the averages of seed-level statistics rather than the statistic of
an ensemble portfolio. Following the delta-method approach of
\citet{ledoit_wolf_2008}, each statistic is written in terms of monthly first
and second moments. The influence functions are averaged across seeds for
each model and then differenced across models, and their Bartlett long-run
variance gives the standard error. IPCA is deterministic and contributes a
single return series. This calculation retains the dependence between models
evaluated on the same months; the seeds are not independent market
histories.

For each criterion and table, the fifteen two-sided $p$-values are adjusted
by Holm's step-down procedure \citep{holm_simple_1979}, which controls the
familywise error rate at $5\%$ within that criterion and table, but not
across all analyses in the paper. \Cref{tab:test_statistics,tab:rolling_tests}
report the statistics. The sign gives the direction of a difference, and
bold statistics have Holm-adjusted $p<0.05$. An insignificant difference
means only that the data cannot order the two models on that criterion.

\begin{table}[!tb]
\centering
\caption{Pairwise Newey--West statistics for the six rows of
\cref{tab:focused_kernels}. Positive values favor the first model; for GMV
this means lower volatility. Bold statistics have Holm-adjusted $p<0.05$,
with fifteen comparisons per criterion. Kernel labels give loadings/factors.}
\label{tab:test_statistics}
\begingroup\small
\setlength{\tabcolsep}{3pt}\renewcommand{\arraystretch}{1.15}
\begin{tabular*}{\textwidth}{@{\extracolsep{\fill}}llrrrrrr@{}}
\toprule
 & & \multicolumn{3}{c}{Return forecasts} & \multicolumn{3}{c}{Portfolio performance} \\
\cmidrule(lr){3-5}\cmidrule(lr){6-8}
First & Second & $R^2$ & Rank & Log score & Tangency & Long--short & GMV vol. \\
\midrule
Affine/const. & Affine/RBF & $-0.36$ & $-1.07$ & $-2.20$ & $-2.01$ & $-0.87$ & $1.78$ \\
Affine/const. & RBF/const. & $-0.07$ & $\boldsymbol{-3.39}$ & $\boldsymbol{-3.31}$ & $-1.24$ & $-0.50$ & $2.29$ \\
Affine/const. & RBF/RBF & $-0.37$ & $-2.34$ & $-2.43$ & $\boldsymbol{-2.87}$ & $-1.00$ & $0.79$ \\
Affine/const. & IPCA & $\boldsymbol{4.23}$ & $2.73$ & $-1.24$ & $\boldsymbol{3.27}$ & $2.01$ & $\boldsymbol{4.46}$ \\
Affine/const. & IPCA+int. & $\boldsymbol{5.38}$ & $1.09$ & $-0.54$ & $1.23$ & $2.01$ & $\boldsymbol{4.79}$ \\
Affine/RBF & RBF/const. & $0.31$ & $-0.70$ & $1.69$ & $-0.75$ & $0.25$ & $1.71$ \\
Affine/RBF & RBF/RBF & $-0.36$ & $\boldsymbol{-3.37}$ & $\boldsymbol{-3.70}$ & $-2.57$ & $-0.53$ & $0.27$ \\
Affine/RBF & IPCA & $1.15$ & $2.78$ & $2.12$ & $\boldsymbol{3.29}$ & $2.48$ & $\boldsymbol{4.12}$ \\
Affine/RBF & IPCA+int. & $1.21$ & $1.57$ & $2.40$ & $1.76$ & $2.04$ & $\boldsymbol{4.60}$ \\
RBF/const. & RBF/RBF & $-0.35$ & $-1.15$ & $-2.00$ & $-2.62$ & $-0.55$ & $\boldsymbol{-4.36}$ \\
RBF/const. & IPCA & $2.56$ & $\boldsymbol{4.01}$ & $0.32$ & $\boldsymbol{2.89}$ & $1.61$ & $2.32$ \\
RBF/const. & IPCA+int. & $\boldsymbol{3.45}$ & $2.78$ & $1.77$ & $2.13$ & $1.46$ & $2.70$ \\
RBF/RBF & IPCA & $0.79$ & $\boldsymbol{3.81}$ & $2.46$ & $\boldsymbol{3.83}$ & $2.38$ & $\boldsymbol{3.06}$ \\
RBF/RBF & IPCA+int. & $0.83$ & $2.75$ & $2.70$ & $\boldsymbol{3.14}$ & $1.86$ & $\boldsymbol{3.33}$ \\
\addlinespace[0.5ex]
IPCA & IPCA+int. & $0.55$ & $-2.62$ & $\boldsymbol{4.01}$ & $-1.72$ & $-1.02$ & $1.95$ \\
\bottomrule
\end{tabular*}
\endgroup

\end{table}

\begin{table}[!tb]
\centering
\caption{Pairwise Newey--West statistics for the six rows of
\cref{tab:rolling}. Signs and adjustment follow \cref{tab:test_statistics}.}
\label{tab:rolling_tests}
\begingroup\small
\setlength{\tabcolsep}{3pt}\renewcommand{\arraystretch}{1.15}
\begin{tabular*}{\textwidth}{@{\extracolsep{\fill}}llrrrrrr@{}}
\toprule
 & & \multicolumn{3}{c}{Return forecasts} & \multicolumn{3}{c}{Portfolio performance} \\
\cmidrule(lr){3-5}\cmidrule(lr){6-8}
First & Second & $R^2$ & Rank & Log score & Tangency & Long--short & GMV vol. \\
\midrule
PanelGP fixed & PanelGP rolling & $1.34$ & $-0.76$ & $-2.55$ & $-2.55$ & $-0.98$ & $\boldsymbol{-3.12}$ \\
PanelGP fixed & IPCA fixed & $0.79$ & $\boldsymbol{3.81}$ & $2.46$ & $\boldsymbol{3.83}$ & $2.38$ & $\boldsymbol{3.06}$ \\
PanelGP fixed & IPCA rolling & $0.21$ & $\boldsymbol{3.17}$ & $0.78$ & $2.07$ & $2.39$ & $-2.09$ \\
PanelGP fixed & IPCA+int. fixed & $0.83$ & $2.75$ & $2.70$ & $\boldsymbol{3.14}$ & $1.86$ & $\boldsymbol{3.33}$ \\
PanelGP fixed & IPCA+int. rolling & $0.31$ & $1.90$ & $0.78$ & $-1.38$ & $0.10$ & $-2.04$ \\
PanelGP rolling & IPCA fixed & $-0.22$ & $\boldsymbol{3.52}$ & $\boldsymbol{3.38}$ & $\boldsymbol{5.21}$ & $2.64$ & $\boldsymbol{4.60}$ \\
PanelGP rolling & IPCA rolling & $-0.98$ & $\boldsymbol{3.14}$ & $\boldsymbol{5.78}$ & $\boldsymbol{3.79}$ & $2.89$ & $2.45$ \\
PanelGP rolling & IPCA+int. fixed & $-0.17$ & $2.70$ & $\boldsymbol{3.48}$ & $\boldsymbol{4.47}$ & $2.28$ & $\boldsymbol{4.94}$ \\
PanelGP rolling & IPCA+int. rolling & $-0.86$ & $2.29$ & $\boldsymbol{5.82}$ & $1.51$ & $0.82$ & $\boldsymbol{2.73}$ \\
\addlinespace[0.5ex]
IPCA fixed & IPCA rolling & $\boldsymbol{-3.16}$ & $-0.10$ & $-1.36$ & $-0.76$ & $0.45$ & $\boldsymbol{-4.33}$ \\
IPCA fixed & IPCA+int. fixed & $0.55$ & $-2.62$ & $\boldsymbol{4.01}$ & $-1.72$ & $-1.02$ & $1.95$ \\
IPCA fixed & IPCA+int. rolling & $-2.55$ & $\boldsymbol{-2.93}$ & $-1.34$ & $\boldsymbol{-4.93}$ & $-2.71$ & $\boldsymbol{-4.25}$ \\
\addlinespace[0.5ex]
IPCA rolling & IPCA+int. fixed & $\boldsymbol{3.16}$ & $-0.97$ & $1.60$ & $-0.69$ & $-1.03$ & $\boldsymbol{4.60}$ \\
IPCA rolling & IPCA+int. rolling & $0.88$ & $-2.16$ & $0.40$ & $\boldsymbol{-2.89}$ & $\boldsymbol{-3.55}$ & $0.20$ \\
\addlinespace[0.5ex]
IPCA+int. fixed & IPCA+int. rolling & $\boldsymbol{-2.97}$ & $-1.76$ & $-1.59$ & $\boldsymbol{-3.92}$ & $-2.42$ & $\boldsymbol{-4.61}$ \\
\bottomrule
\end{tabular*}
\endgroup

\end{table}

\subsection{Results by subperiod}
\label{app:subperiods}

\Cref{tab:subperiods} splits the 360 evaluation months of \cref{tab:rolling}
into two halves of 180 input months, 1995--2009 and 2010--2024. Every model
has lower Sharpe ratios and rank correlations in the second half. The
decline is common to PanelGP and IPCA, in line with the attenuation of
cross-sectional return predictability documented after the early 2000s
\citep{chordia_have_2014,mclean_does_2016}. PanelGP has the highest Sharpe
ratios and rank correlations in both halves. In 2010--2024, the tangency
portfolios of rolling PanelGP and rolling IPCA with an intercept exceed the
Sharpe ratio of the S\&P 500 over the same months, 0.93; the
index is not a market-neutral strategy and serves only as a reference level.
With 180 months per half, the standard error of an annualized Sharpe ratio
is about 0.26 or larger, so differences between models within a half are
imprecisely estimated.

\begin{table}[!tb]
\centering
\caption{\Cref{tab:rolling} by half of input months (1995--2009 and
2010--2024), $H=5$. PanelGP uses RBF loadings and RBF factors, and its
entries average five seeds. Sharpe ratios are annualized and before trading
costs. Rank is the average monthly Spearman correlation. The S\&P 500 has
Sharpe ratios of 0.34 and 0.93 over the
realization months of the two halves.}
\label{tab:subperiods}
\begingroup\small
\setlength{\tabcolsep}{4pt}\renewcommand{\arraystretch}{1.15}
\begin{tabular*}{\textwidth}{@{\extracolsep{\fill}}llrrrrrr@{}}
\toprule
 & & \multicolumn{2}{c}{Tangency Sharpe} & \multicolumn{2}{c}{Long--short Sharpe} & \multicolumn{2}{c}{Rank (\%)} \\
\cmidrule(lr){3-4}\cmidrule(lr){5-6}\cmidrule(lr){7-8}
Model & Training & 95--09 & 10--24 & 95--09 & 10--24 & 95--09 & 10--24 \\
\midrule
PanelGP & Fixed & 2.31 & 0.74 & 1.34 & 0.53 & 6.80 & 2.24 \\
PanelGP & Rolling & 2.62 & 1.16 & 1.36 & 0.78 & 6.16 & 3.53 \\
\addlinespace[0.5ex]
IPCA & Fixed & 1.22 & 0.38 & 0.71 & 0.39 & 3.68 & 1.50 \\
IPCA & Rolling & 1.11 & 0.92 & 0.66 & 0.31 & 3.99 & 1.29 \\
\addlinespace[0.5ex]
IPCA+int. & Fixed & 1.74 & 0.52 & 0.90 & 0.34 & 4.24 & 1.98 \\
IPCA+int. & Rolling & 2.40 & 1.04 & 1.29 & 0.54 & 5.22 & 2.16 \\
\bottomrule
\end{tabular*}
\endgroup

\end{table}

\subsection{Log-score gain of the bilinear predictive density}
\label{app:density_gain}

\begin{figure}[!tb]
\centering
\includegraphics[width=\textwidth]{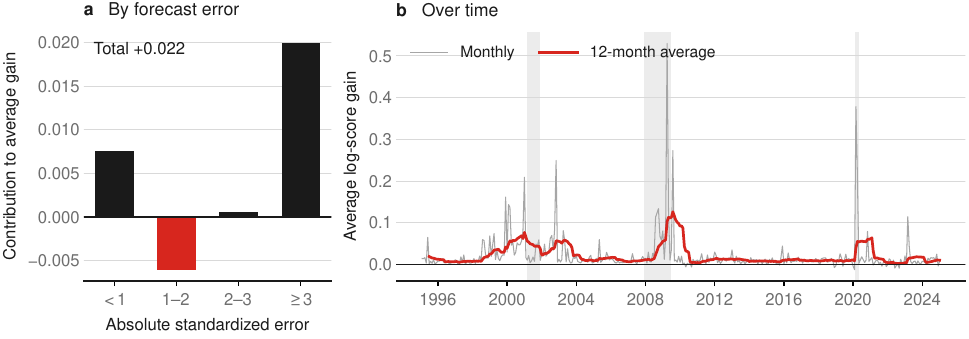}
\caption{Log-score gain $g_{it}$ of the bilinear predictive density over the
Gaussian density with the same mean and variance, under annual re-estimation.
Left: contribution of each group of absolute standardized errors to the
average gain; the four bars sum to the average gain. Right: monthly average
gain and its trailing 12-month average; shading marks NBER peak-to-trough
intervals. Dates are return realization months. Results pool five seeds.}
\label{fig:density_gain}
\end{figure}

\Cref{fig:density_gain} shows where the log-score gain $g_{it}$ of
\cref{eq:bilinear_density_gain} comes from, by size of the standardized
forecast error and over time. The density is computed as in
\cref{app:marginal_density}.

\FloatBarrier

\end{document}